\documentclass[a4paper,10pt]{article}
\usepackage{graphicx} 
\usepackage{graphics} 
\usepackage{amsmath}
\usepackage{hyperref}

\newcommand{\ket}[1]{|#1\rangle}
\newcommand{\bra}[1]{\langle#1|}
\newcommand{\defi}[6]{\begin{large}\texttt{#1}(\textbf{#2})\end{large}:
\small{#3} \\ \hskip 0.5cm [ {\sf Input}: #4, {\sf Output}: #5, {\sf Calls}: #6]}  
\newcommand{\Id}[0]{1\!\!\:\!{\rm{I}}  }
\newcommand{\trace}[2]{\mathrm{Tr}_{#1}\left\{#2\right\}}
\title{\textsf{Quantavo}:  a Maple Toolbox for Linear Quantum Optics}

\author{$^*$Alvaro Feito Boirac$^{1,2}$\\
\\
\small{$^1$Institute for Mathematical Sciences, Imperial College London,}\\ 
\small{53 Exhibition Road, London SW7 2PG, UK}\\
\small{$^2$QOLS, The Blackett Lab. Imperial College London,}\\
\small{Prince Consort Road, London SW7 2BW, UK}}

\date{\today}

\begin{document}
\maketitle
\begin{abstract}
This manual describes the basic objectives, functionalities and uses of the 
toolbox for \textit{Maple} (Maplesoft$^{TM}$) called \textsf{Quantavo}.  It is intended to facilitate calculations both symbolically and numerically related to Quantum Optics.  In particular the evolution, measurement and entanglement properties of quantum states in the Fock basis can be simulated with it.  It is provided to the community as a free open source module.
\end{abstract}
\vfill
*\href{mailto:ab1805@imperial.ac.uk}{ab1805@imperial.ac.uk}
\newpage
\tableofcontents

\section{Objectives}

The toolbox is made to be used with the following approach:\\\\
Declare an initial state, let it evolve through a quantum optical circuit involving 
linear optics (LO) and measurements and finally ask various questions about the structure, entanglement,
and properties of the final state.\\\\

The toolbox intends to do so providing:\\\\
A) A framework in which to declare, manipulate and characterize quantum states
of light (finite number of modes, and finite dimensional).\\\\
B) Procedures that implement linear operations or whole linear optics circuits on our states such as:
\begin{itemize}
\item Beam Splitters (BS) 
\item Phase Shifters (PS).
\item arbitrary unitary transformations of the modes.
\end{itemize}
C) Procedures that implement arbitrary measurements (both projective
or generalized positive operator valued measures (POVM)).\\\\
D) Procedures to determine probabilities and expected values for projective measurements
and POVM measurements.\\\\
E) Procedures to trace out measured or inaccessible modes.\\\\
F) Procedures to calculate different entanglement measures such as:
\begin{itemize}
\item Entropy of entanglement
\item partial trace, norm  $\rightarrow$ Negativity, Logarithmic Negativity
\end{itemize}
G) Access to properties such as the mean Energy of given states.\\\\
H) Extract and display lists of coefficients along with their indexes like: $RT\sqrt{2}$  $|001\rangle \langle001|  $
\\
\\
H) Tools to easily plot states and density matrices.

\section{Getting Started}
Some previous knowledge about \textit{Maple} from Maplesoft$^{TM}$ is required to use this tool.
However, the brief ``Take a Tour of \textit{Maple}'' should suffice to get started.  This toolbox works
with Maple 9.5, Maple 10 and Maple 11.
\subsection{Definitions and Notation}
\textbf{Procedures}:  Formally, in Maple, a procedure definition is a valid expression that can be assigned to a name.  The procedures we will use can be thought of as a set of ``rules'' that generally receive one or more inputs and return one or more outputs.  We will write them in typewriter face.  For example the procedure \texttt{IsHermitian}(M) evaluates if a Matrix \textbf{M} describing a density matrix is hermitian. A dictionary with all the procedures involved in \textbf{Quantavo} can be found in the appendix.
\\
\\
\textbf{Modules}:  Modules are repositories of procedures.  By loading a module we can use its procedures in the
Maple worksheet.  For example \textbf{Quantavo} is the module containing the procedures we will use.
\\
\\
\textbf{objects}:  The objects in which we will encode our quantum states will be written in bold
face.  These will include \textbf{vec, mat, matcol} and \textbf{poly} and will be introduced later.
\\
\\
\textbf{d and K}:  Throughout the manual, `$K$' will stand for the number of modes and `$d$' for the
dimension of each mode.  If considering the photon number degree of freedom, then  `$d-1$' will be the maximum number of photons in any mode.  It is important to keep track of the value of these two
global variables throughout the worksheet as they play an important role in the labelling of the optical modes and translation procedures.  These variables can always be updated and displayed with the procedure:  \texttt{findKnd}(State).
\subsection{Loading the Modules}
All necessary files can be found at,\\
\url{http://www.imperial.ac.uk/quantuminformation/research/downloads}
\begin{enumerate}
 \item uncompress the file \textbf{Quantavo.zip}.
 \item Save the folder QUANTAVO to a given directory.  It should contain the files
 \textbf{Quantavo.mpl} and \textbf{Quantavo\_Example\_Worksheet.mw}.
 \item One can start opening the worksheet \textbf{Quantavo\_Example\_Worksheet.mw}.
 \item To use the module in a new worksheet execute the following commands \footnote[1]{note that if the new file is in a different directory it should be,\\
$>$read ``/path-to-folder/QUANTAVO/Quantavo.mpl'';\\
or under MS Windows,\\
$>$read ``C:$\backslash$$\backslash$path-to-folder$\backslash$$\backslash$QUANTAVO$\backslash$$\backslash$Quantavo.mpl''; }:
\\
\\
 \frame{
\begin{tabular}{lc} \\
&   
\begin{minipage}{4in}
$>$ with(LinearAlgebra):\\
$>$ read ``Quantavo.mpl'';\\
$>$ with(Quantavo);\\
\\
\end{minipage}
\end{tabular}
}
\\
\\
It should return a list of all the procedures available \footnotemark[1]:
\\
\\
 \frame{
\begin{tabular}{lc} \\
&   
\begin{minipage}{4.5in}
 [APD, BS, BuildUnitary, CoherentState, DP, Dbra, Dbraket, DeltaK, Dket, Dstate, Energy, Entropy, EvalState, IdentityState, IsHermitian, IsNormalized, LogNegativity, Negativity, POVMresult, PS, PlotState, Probability, Project, SqueezedVac, StateApprox, StateComplexConjugate, StateMultiply, StateNorm, StateNormalize, StatePartialTranspose, StateSort, StateTrace, TensorProduct, TensorVac, Traceout, Trim, UnitaryEvolution, Vac, findKnd, indexstate, mat2matcol, mat2poly, matcol2mat, matcol2poly, modesmatcol, myBS, poly2matcol, poly2vec, vec2mat, vec2matcol, vec2poly]
\\
\\
\end{minipage}
\end{tabular}
}
\item  You are ready to use \textsf{Quantavo} !\\
\end{enumerate}

Note that You may also save the module to your Maple library.  To do so,
visit the maple help on \texttt{module}, \texttt{savelibname} and \texttt{savelib}.

\newpage

\section{Toolbox}
To run \textsf{Quantavo} one has to load the following modules:
\begin{itemize}
\item \textbf{LinearAlgebra} (Linear Algebra package from Maplesoft$^{TM}$ built in \textit{Maple})
 \item \textbf{Quantavo}  (General toolbox)
   \end{itemize}
Additionally, if we want to plot our states we will also need the module
\begin{itemize}
\item \textbf{geom3d} (geometry package from Maplesoft$^{TM}$ built in \textit{Maple})
\item \textbf{plots} (plotting package from Maplesoft$^{TM}$ built in \textit{Maple})
 \end{itemize}

\subsection{Objects and Operations}

When quantum optical states have a few modes and live in high dimensions, the matrices or vectors describing them
very soon become intractable.  To mitigate this difficulty, the procedures from \textbf{Quantavo} store and manipulate only the non-zero elements in the description of our states.  Our main objects will be 2 column and 3 column matrices.

\subsubsection{Pure States}
Two column objects will describe vectors in Hilbert space of the form:
$$\ket{\phi}= \sum_{n_1,n_2,...,n_K}^{d-1} f(n_1,n_2,...,n_K) \ket{n_1,n_2,...,n_K} $$
Where all indices ``$n_i$'' range from $0 \rightarrow (d-1)$.
Pure state vectors will be encoded in 2 column matrices that we will call \textit{trimmed vectors} containing only non-zero entries.  They will only contain non-zero entries.  These objects will be named as a short vector:  ``\textbf{vec}''
and will have the following appearance:

$$\psi := \left[ \begin {array}{cc} 1&[0,0,0]\\\noalign{\medskip}\lambda&[1,1,0
]\\\noalign{\medskip}{\lambda}^{2}&[2,2,0]\\\noalign{\medskip}{\lambda
}^{3}&[3,3,0]\end {array} \right] $$

For each row,
the second column will be a list with the number of photons in each mode,  This way [0,1,2] stands for $\ket{012}$.
the first column will contain the coefficient associated with this ket.
The whole will describe the linear supperposition of all these kets with their coefficients, therefore the above $\psi$ describes the unnormalized quantum state:
$$\ket{\psi} = \sum_{n=0}^{3} \lambda^n \ket{n,n}.$$
\\\\

\subsubsection{Mixed States}

Two objects will be used to display density matrices.  The first one is a square matrix, with as little
zero entries as possible.  This matrix will be called ``\textbf{mat}'' and will have for the above state the following
form:

 $$\rho = \left[ \begin {array}{ccccc} 0&[0,0,0]&[1,1,0]&[2,2,0]&[3,3,0]
\\\noalign{\medskip}[0,0,0]&1&\bar{\lambda} &
  \bar{\lambda}^{2}& \bar{\lambda}^{3}
\\\noalign{\medskip}[1,1,0]&\lambda&\lambda\,\bar{\lambda} &\lambda\,  \bar{\lambda}^{2}&\lambda\, \bar{\lambda}^{3}\\\noalign{\medskip}[2,2,0]&{\lambda}^{2}&{
\lambda}^{2}\bar{\lambda} &{\lambda}^{2}
  \bar{\lambda}   ^{2}&{\lambda}
^{2} \bar{\lambda}^{3}
\\\noalign{\medskip}[3,3,0]&{\lambda}^{3}&{\lambda}^{3}\bar{\lambda}  &{\lambda}^{3} \bar{\lambda}^{2}&{\lambda}^{3} 
\bar{\lambda}^{3}\end {array} \right]   $$\\\\

Another object that can also describe a density matrix will be a 3 column matrix, or trimmed density matrix. It will be named as a short column matrix:  ``\textbf{matcol}''.
The second column will be a list with the number of photons in each mode of the ket.  This way,  [0,1] in the 2nd column describes $\ket{01}$.  The third column will be a list with the number of photons in each mode of the bra; therefore [2,1] in the 3rd column describes $\bra{21}$. Finally The first column will have the coefficient associated with this $|$ket$\rangle \langle$bra$|$ .
The whole will describe the non zero elements of the density matrix.  As an example consider the above state which will be:

$$\rho_{matcol} = \left[ \begin {array}{ccc} 1&[0,0,0]&[0,0,0]\\\noalign{\medskip}
\bar{\lambda} &[0,0,0]&[1,1,0]\\\noalign{\medskip}
 \bar{\lambda}^{2}&[0,0,0]&[2,2,0]\\\noalign{\medskip} 
 \bar{\lambda}^{3}&[0,0,0]&[3,3,0]\\\noalign{\medskip}
 \lambda&[1,1,0]&[0,0,0]\\\noalign{\medskip}
 \lambda\,\bar{\lambda} &[1,1,0]&[1,1,0]\\\noalign{\medskip}\lambda\, 
\bar{\lambda}^{2}&[1,1,0]&[2,2,0]\\\noalign{\medskip} \lambda\, \bar{\lambda}^{3}&[1,1,0]&[3,3,0]\\\noalign{\medskip}
{\lambda}^{2}&[2,2,0]&[0,0,0]\\\noalign{\medskip}{\lambda}^{2}\bar{\lambda}
 &[2,2,0]&[1,1,0]\\\noalign{\medskip}{\lambda}^
{2}  \bar{\lambda}^{2}&[2,2,0
]&[2,2,0]\\\noalign{\medskip}{\lambda}^{2} 
\bar{\lambda }^{3}&[2,2,0]&[3,3,0]
\\\noalign{\medskip}{\lambda}^{3}&[3,3,0]&[0,0,0]\\\noalign{\medskip}{
\lambda}^{3}\bar{\lambda} &[3,3,0]&[1,1,0]
\\\noalign{\medskip}{\lambda}^{3}  \bar{\lambda}^{2}&[3,3,0]&[2,2,0]\\\noalign{\medskip}{
\lambda}^{3}  \bar{ \lambda }^{3}&[3,3,0]&[3,3,0]\end {array} \right] $$

Summarising, we will use mainly 3 objects:  \textbf{vec}, \textbf{mat} and \textbf{matcol}.\\\\
A fourth object less commonly used is a polynomial representation of the state.  It is a polynomial in the mode operators that define the state.  A general state would then be:
$$\rho= \sum_{\begin{array}{c}
n_1,n_2,...,n_K\\
m_1,m_2,...,m_K
\end{array}}^{d-1} f(n_1,n_2,...,n_K,m_1,...,m_K)\;\; a_1^{\dagger n_1} a_2^{\dagger n_2} ... a_K^{\dagger n_K} \; \ket{0} \bra{0} \; b_1^{m_1} b_2^{m_2} ... b_K^{m_K} $$
and its description as a \textbf{poly} object:
$$poly(\rho) =  \sum_{\begin{array}{c}
n_1,n_2,...,n_K\\
m_1,m_2,...,m_K
\end{array}}^{d-1} f(n_1,n_2,...,n_K,m_1,...,m_K)\;\; a_1^{n_1} a_2^{n_2} ... a_K^{n_K} \;  \; b_1^{ m_1} b_2^{ m_2} ... b_K^{ m_K} $$
(note that the commutation relations are not taken care of in \textbf{poly} objects).

\subsection{Declare, Propagate, Measure and Ask}

\textbf{Quantavo} contains various procedures that allow problems to be formulated as follows:
\begin{enumerate}
 \item Declare the initial state.
 \item Apply different transformations to it (Beam Splitter, Phase Shifter, Arbitrary Unitary, ...).
 \item Measure certain modes (and trace out the inaccessible ones), find out probabilities.
 \item Ask different questions about the properties of the state:  Display, Plot, evaluate certain measures of Entanglement, etc..
\end{enumerate}

Additionally, the order in which we use these procedures can be changed and the questions in item (4) can be formulated at any intermediate time.  In addition, more states can be tensored or added at later times.  Finally there are procedures to interconvert \textbf{mat}, \textbf{vec}, \textbf{matcol} and \textbf{poly}.  Let us then give a more detailed
description of these four basic steps.

\subsubsection{Declaration}

There are different ways to declare a state, all depending on its characteristics.\\\\
\textbf{Pure States}\\\\
For pure states of known functional form.  That is, if we have a state of the form:
\begin{equation}\label{pure}
\ket{\psi}= \sum_{n_1,n_2,...,n_K}^{d} f(n_1,n_2,...,n_K) \ket{n_1,n_2,...,n_K}
\end{equation}and we know explicitly $f(n_1,n_2,...,n_K)$ we may use the following structure:\\\\
Declare the number of modes ``$K$'', the maximum number of photons ``$d$'', and make a loop to declare the elements.  For example:\\\\
\frame{
\begin{tabular}{cc} \\
&   
\begin{minipage}{4in}
 d := 4;\\
 K := 3; \\
 V:=Matrix$(d^K,2)$:\\
\end{minipage}\\
\end{tabular}
}
\\
There are now $2 d^K$ elements to be specified.  
\\
\\
\frame{
\begin{tabular}{cc} \\
&   
\begin{minipage}{4in}
for i from 1 to $d$ do
\\
for j from 1 to $d$ do
\\
for k from 1 to $d$ do \\\\
V[i,1]:=$f$(i,j,k):\\
V[i,2] :=[i,j,k]:\\\\
end do: \\
end do: \\
end do: \\
\\
\end{minipage}
\end{tabular}
}
\\\\
Where $f$ is the function in eq.(\ref{pure}).  We may use the
procedure \texttt{deltaK}(i,j) if we need a Kronecker delta in our definition.
Executing the above loop will declare a matrix \textbf{V} of size $d^K \times 2$ 
that has hopefully many zero entries.  
To get rid of the zero entries and convert this object 
into a \textbf{vec} object (or trimmed vector) we will use the 
procedure \texttt{Trim}: \\\\
\frame{
\begin{tabular}{cc} \\
&   
\begin{minipage}{4in}
V1:=\texttt{Trim}(V):\\
\end{minipage}
\end{tabular}
}\\\\
Another way to proceed is to declare the object \textbf{vec} directly with an appropriate function.
Consider as an example the state:
$$\ket{\psi}=\sum_{n=0}^{3} \lambda^n \ket{n,n,0}$$
In this case it is easy to declare the object \textbf{vec} directly as:\\
\frame{
\begin{tabular}{cc} \\
&   
\begin{minipage}{4in}
V1:=Matrix(4,2):\\
for i from 1 to 4 do\\
V1[i,1]:=$\lambda^{i-1}$:\\
V1[i,2]:=[i-1,i-1,0]:\\
end do:\\
\end{minipage}
\end{tabular}
}\\\\
However this is not allways the case and if we have a pure state with no known $f(n_1,n_2,...,n_K)$, but we know
which non-zero elements it contains, we can declare its elements one by one or declare the \textbf{vec} matrix at once.  For example to declare $$\ket{\psi} = \ket{00}+\lambda\ket{11}+\lambda^2\ket{21}$$ we can use:
\\\\
\frame{
\begin{tabular}{cc} \\
&   
\begin{minipage}{4in}
d:=2:\\
K:=2:\\
V:=Matrix(3,2):\\
V[1,1]:=1:\\
V[1,2]:=[0,0]:\\
V[2,1]:=$\lambda$:\\
V[2,2]:=[1,1]:\\
V[3,1]:=$\lambda^2$:\\
V[3,2]:=[2,1]:\\
\end{minipage}
\end{tabular}
}\\\\
or\\\\
\frame{
\begin{tabular}{cc} \\
&   
\begin{minipage}{4in}
d:=2:\\
K:=2:\\
V:=Matrix([ [1,[0,0]],
[$\lambda$,[1,1]],
[$\lambda^2$,[2,1]]
]):
\end{minipage}
\end{tabular}
}\\
\\
or\\\\
\frame{
\begin{tabular}{cc} \\
&   
\begin{minipage}{4in}
d:=2:\\
K:=2:\\
V:=$<<$(1, $\lambda$, $\lambda^2$)$>|<$[0, 0], [1, 1], [2, 1]$>>$:\\
\end{minipage}
\end{tabular}
}\\
Adding \texttt{StateNormalize(V)} for normalization.
\\
\\
A special family of pure states are readily available in {\sf Quantavo}:
\\
\\
\textbf{Squeezed vacuum}:\\
for a truncated, unnormalized, pure single mode squeezed vacuum state $\ket{\phi} \sim \sum_{n=0}^{d-1} \lambda^n \ket{n}$ type:
\\
\\
\frame{
\begin{tabular}{cc} \\
&   
\begin{minipage}{4in}
\texttt{SqueezedVac}(1, d, $\lambda$);\\
\end{minipage}
\end{tabular}
}\\
\\
for a truncated, unnormalized, pure two mode squeezed vacuum state\\ $\ket{\phi} \sim \sum_{n=0}^{d-1} \lambda^n \ket{n,n}$ type:
\\
\\
\frame{
\begin{tabular}{cc} \\
&   
\begin{minipage}{4in}
\texttt{SqueezedVac}(2, d, $\lambda$);\\
\end{minipage}
\end{tabular}
}
\\
\\
These states are given without normalization, since for finite $d$, $\sqrt{1-\lambda^2}$ doesn't normalize them.  Calculations and displays are easier this way.
\\
\\
\textbf{Coherent States}:\\
for a truncated single mode coherent state $\ket{\phi} = \sum_{n=0}^{d-1} \frac{\alpha^n}{\sqrt{n!}} \ket{n}$ type:\\
\\
\frame{
\begin{tabular}{cc} \\
&   
\begin{minipage}{4in}
\texttt{CoherentState}(1, d, $\alpha$);\\
\end{minipage}
\end{tabular}
}\\
\\
The analytical normalization is rather lengthy so it is left unnormalized.
Normalization can be done at a later time with \texttt{StateNormalize}.
\\
\\
It is also possible to tensor some vacuum modes to our state.  The procedure \texttt{TensorVac}($\psi$, s)  effectively does the following transformation:  $\ket{\psi}  \longrightarrow  \ket{\psi}\otimes\ket{0}^{\otimes s}$
when applied either to a \textbf{vec}, \textbf{mat} or \textbf{matcol}  (see appendix for more details).\\\\
\\
\\
\textbf{Tensor Product}:\\
The procedure \texttt{TensorProduct}(A, $List_A$, B, $List_B$) will make the tensor product between modes [$List_A$] of state A and modes [$List_B$] of state B.  So for example,
\\
\frame{
\begin{tabular}{cc} \\
&   
\begin{minipage}{4in}
Co:=\texttt{CoherentState}(1, 4, $\alpha$);\\
Fock:=Matrix([1,[1]]);\\
State:=TensorProduct(Co,[1],Fock,[2]);
\\
\\
\end{minipage}
\end{tabular}
}\\
\\
will result in the following \textbf{vec} object,
$$\mathrm{State} = \left[ \begin {array}{cc} 1&[0,1]\\\noalign{\medskip}\alpha&[1,1]
\\\noalign{\medskip}1/2\,{\alpha}^{2}\sqrt {2}&[2,1]
\\\noalign{\medskip}1/6\,{\alpha}^{3}\sqrt {6}&[3,1]\end {array}
 \right] 
$$

\textbf{Mixed States}\\

The generalized states (either pure or mixed) we are interested in can be written as:
\begin{equation}\label{mixed}
\rho= \sum_{\widehat{n},\widehat{m}}^{d-1} g(\widehat{n},\widehat{m}) \ket{\widehat{n}}\bra{\widehat{m}}
\end{equation}
where $\widehat{n}$ and $\widehat{m}$ stand for $n_1,n_2,...,n_K$ and $m_1,m_2,...,m_K$ respectively.
The density matrix for these states is $d^K \times d^K$ dimensional which is in general
too large for the computer to handle.  We will therefore describe these states using \textbf{mat}, \textbf{matcol} or \textbf{poly}.  
objects.\\\\
Two main strategies can be used to declare our states:
\begin{enumerate}
 \item If our starting state is pure and will become mixed later, we can declare a pure \textbf{vec} object and then convert it into a \textbf{mat} or \textbf{matcol} when needed.
 All procedures to convert are named in the intuitive way: ``\texttt{object2object}''.  This way to convert
 \textbf{vec} into \textbf{mat} we have the procedure \texttt{vec2mat}, to convert \textbf{mat} into \textbf{matcol}, \texttt{mat2matcol} and so on for all objects and conversions.
 Therefore, once our pure state vector \textbf{V} has been declared we can do the following:\\
 \\
\frame{ 
\begin{tabular}{cc} \\ 
&    
\begin{minipage}{4in} 
V1:=\texttt{Trim}(V):  \#eliminate non-zero entries\\
M1:=\texttt{vec2mat}(V1):  \#convert it to a density matrix\\       
\end{minipage} 
\end{tabular} 
}\\
or\\\\
\frame{ 
\begin{tabular}{cc} \\ 
&    
\begin{minipage}{4in}  
V1:=\texttt{Trim}(V): \\ 
M1col:=vec2matcol(M1):\\    
 \end{minipage}  
\end{tabular}  
}
\item if our starting state is mixed to begin with, we can declare our initial state as the object
\textbf{mat} or \textbf{matcol}.  To do so, if we know the functional form of $g(\widehat{n},\widehat{m})$ in eq.(\ref{mixed}) then we may directly declare our state.  For example to declare the state:
$$\rho = \sum_{n,m=0}^{4} \lambda^{n+m} \ket{n,n,0}\bra{m,m,0}$$
We could use:

\frame{ 
\begin{tabular}{cc} \\ 
&    
\begin{minipage}{4in}  
M1col:=Matrix(3,25): \\ 
for i from 1 to 5 do\\
for j from 1 to 5 do\\

M1col[i,1]:=$\lambda^{i+j-2}$;\\
M1col[i,2]:=[i-1,i-1,0];\\
M1col[i,3]:=[j-1,j-1,0];\\\\
end do:\\
end do:
\\
\end{minipage}  
\end{tabular}  
}

or

\frame{ 
\begin{tabular}{cc} \\ 
&    
\begin{minipage}{4in}  
V1:=SqueezedVac(2, 4, lambda): \\ 
V1:=TensorVac(V1, 1):\\
M1col:=vec2matcol(V1):\\    
 \end{minipage}  
\end{tabular}  
}

or yet again, meaning 
$\rho = \left[\left(\sum_{n=0}^{4} \lambda^n \ket{n} \right) \otimes \left( \sum_{m=0}^{4} \lambda ^m \ket{m}\ket{0} \right) \right] \left[ c.c. \right]$

\frame{ 
\begin{tabular}{cc} \\ 
&    
\begin{minipage}{4in}  
V1:=SqueezedVac(1, 5, lambda): \\ 
V2:=SqueezedVac(1, 5, lambda): \\ 
V2:=TensorVac(V2, 1):\\
V3:=TensorProduct(V1,[1],V2,[2,3]);\\
M1col:=vec2matcol(V3):\\    
 \end{minipage}  
\end{tabular}  
}

\end{enumerate}

The basic conclusion is that there is no single way of declaring our state
and that depending on its structure we have to find a clever way of declaring it.
As a general guideline, small states without an obvious functional structure can be declared 
giving all elements.  Medium sized pure states can be declared using a clever loop, `trimmed'
and transformed to \textbf{vec}, \textbf{mat} or \textbf{matcol}.
\\
\\
It is worth noting that when using the built-in declaration procedures such as
\texttt{SqueezedVac}, \texttt{CoherentState}, \texttt{TensorVac}, \texttt{Vac},
or the inter-converting ones like \texttt{mat2matcol}, \texttt{mat2poly}, \texttt{vec2mat}, etc, the values of $K$ and $d$ are automatically updated.  However, if the states are declared
from scratch, the values of $K$ and $d$ should be explicitly declared.  Applying to our
state $\rho$ the procedure \texttt{findKnd}($\rho$) can help troubleshoot by reevaluating the value of these global variables.

\subsubsection{Evolution}

\textbf{Beam Splitter}\\\\
We can apply a beam splitter to a \textbf{vec} or \textbf{matcol} state \textbf{V} with the \texttt{BS} procedure:
\\
\\
\frame{ 
\begin{tabular}{cc} \\ 
&    
\begin{minipage}{4in}  
V1:= \texttt{BS}( V , i , j ):
\\
\end{minipage}  
\end{tabular}  
}

Where i, j specify which modes the beam splitter acts on.  It is therefore essential to carefully label
the modes of our quantum optical circuit.  This will effectively do the following mode transformation:

\begin{equation}
\left(
\begin{array}{c}
 a_i'\\ a_j'
\end{array} \right) 
=
\left( 
\begin{array}{cc}
t & r\\
-r & t
\end{array}
\right)
\left(
\begin{array}{c}
 a_i\\ a_j
\end{array} \right) 
\end{equation}\\\\
Leaving `$t$' and `$r$' as unevaluated variables.  If we want to use different reflectivities and transmittivities for our Beam Splitter (variables or numbers) we may use \texttt{myBS} and input  `$t=t_0$' and `$r=r_0$' as follows:
\\
\\
\frame{ 
\begin{tabular}{cc} \\ 
&    
\begin{minipage}{4in}  
V1:= \texttt{myBS}( V , i , j , $t_0$ , $r_0$ ):
\\
\end{minipage}  
\end{tabular}  
}
\\
For the mixed state object \textbf{matcol} we may use the same procedures and the effective transformation will be:

\begin{eqnarray}
\left(
\begin{array}{c}
 a_i'\\ a_j'
\end{array} \right) 
=
\left( 
\begin{array}{cc}
t & r\\
-r & t
\end{array}
\right)
\left(
\begin{array}{c}
 a_i\\ a_j
\end{array} \right) \\
\left(
\begin{array}{c}
 b_i'\\ b_j'
\end{array} \right) 
=
\left( 
\begin{array}{cc}
t^* & r^*\\
-r^* & t^*
\end{array}
\right)
\left(
\begin{array}{c}
 b_i\\ b_j
\end{array} \right). 
\end{eqnarray}
Further details about its use can be found in the example in section (\ref{example})
\\
\\
\textbf{Phase Shifter}\\\\
The phase shifter procedure (PS) can be used with mixed and pure states:\\

\frame{ 
\begin{tabular}{cc} \\ 
&    
\begin{minipage}{4in}  
V1:= \texttt{PS}( \textbf{vec}/\textbf{matcol} , i ,$\phi$ ):
\\
\end{minipage}  
\end{tabular}  
}
\\
The input \textbf{vec}/\textbf{matcol} means that either \textbf{vec} or \textbf{matcol} objects can be given as inputs.  This
procedure makes the effective transformation:
\begin{eqnarray}
a_i' = e^{i \phi} a_i\\
b_i' = e^{-i \phi} b_i
\end{eqnarray}
\\
\\
\textbf{Build Unitary}\\\\
If we wish to construct a unitary matrix that transforms the modes of light and describes a given linear optics (LO) circuit we may use the procedure \texttt{BuildUnitary}.  Together with \texttt{UnitaryEvolution} it will evolve the state through a given LO circuit.  To build a unitary consisting of Beam Splitters (BS) and Phase Shifters (PS) we will do the following:
\\
\\
\frame{ 
\begin{tabular}{cc} \\ 
&    
\begin{minipage}{4in}  
U:= \texttt{BuildUnitary}([List]):
\\
\end{minipage}  
\end{tabular}  
}
\\
\\
This will create a matrix \textbf{U} of dimension $K \times K$, that later will transform the $K$ modes.
The list that is \texttt{BuildUnitary}'s input must have a precise format.  It must be a list of lists.
For example:
List:=[[1,2,t,r],[3,phi],[3,4,q]]; \\
means that first a BS with transmitivity $t^2$ and reflectivity $r^2$ will be applied to modes 1 and 2, then a PS will be applied to mode 3 and finally a BS with transmitivity $q^2$ and reflectivity $(1-q^2)$ will be the last operation.  If we have 4 modes, this will build the matrix
\begin{eqnarray*} U & = &\left( \begin {array}{cccc} t&r&0&0\\\noalign{\medskip}-r&t&0&0
\\\noalign{\medskip}0&0&{e^{i\phi}}q&{e^{i\phi}}\sqrt {1-{q}^{2}}
\\\noalign{\medskip}0&0&-\sqrt {1-{q}^{2}}&q\end {array} \right)  \\
 &=&\left( \begin {array}{cccc} 1&0&0&0\\\noalign{\medskip}0&1&0&0
\\\noalign{\medskip}0&0&q&\sqrt {1-{q}^{2}}\\\noalign{\medskip}0&0&-
\sqrt {1-{q}^{2}}&q\end {array} \right)
  \left( \begin {array}{cccc} 1&0&0&0\\\noalign{\medskip}0&1&0&0
\\\noalign{\medskip}0&0&{e^{i\phi}}&0\\\noalign{\medskip}0&0&0&1
\end {array} \right)
 \left( \begin {array}{cccc} t&r&0&0\\\noalign{\medskip}-r&t&0&0
\\\noalign{\medskip}0&0&1&0\\\noalign{\medskip}0&0&0&1\end {array}
 \right) 
\end{eqnarray*}
\\
\\
In general, in our list of lists, lists with 4 elements, like [i, j, t, r] build BS transformations between
modes i and j, lists with 3 elements like [i, j, t] build BS transformations for modes $i$ and $j$ such that $r=\sqrt{1-t^2}$, and lists
with two elements like [i, $\phi$] build PS transformations on mode i.
\\
\\
\textbf{UnitaryEvolution}\\\\
Whether we have just built a unitary matrix with \texttt{BuildUnitary} or we have a $K \times K$ arbitrary unitary matrix to 
transform our modes $\{a_i\}$, we can use this procedure as follows:
\\
\\
\frame{ 
\begin{tabular}{cc} \\ 
&    
\begin{minipage}{4in}  
V:= \texttt{UnitaryEvolution}(U, \textbf{vec/matcol}): 
\\
\end{minipage}  
\end{tabular}  
}
\\
\\
where U is the unitary matrix of dimension $K \times K$ and our state is described by a
\textbf{vec} or a \textbf{matcol} object.  This will effectively implement the mode transformation:

$$ \overline{a}' = U \overline{a}$$
$$ \overline{b}'^\dagger = U^\dagger \overline{b}^\dagger$$

\subsubsection{Measurement}

\hskip 0.55cm \textbf{The} \texttt{Project}  \textbf{Procedure}:\\\\
If we wish to know the state after a projective measurement we may use the procedure
\texttt{Project}.  Depending on the inputs we give to the procedure it will
do a projective measurement and return a density matrix (\textbf{matcol}) or state vector (\textbf{vec}).  Below is a
description of its different uses:

\begin{enumerate}
\item if given (\textbf{vec$_1$}, list, \textbf{vec$_2$}) and say \textbf{vec$_1$} and \textbf{vec$_2$} describe respectively $\ket{\psi_1}$ and $\ket{\psi_2}$, \texttt{Project}(\textbf{vec$_1$}, list, \textbf{vec$_2$}) returns the \textbf{vec}
(in principle unnormalized) corresponding to the expresion:
$$\ket{\psi_2'} = \left(\ket{\psi_1}\bra{\psi_1}_{list} \otimes 1\!\!\:\!{\rm{I}}_{rest}\right) \;\; \ket{\psi_2}$$
As an example consider list=[2,3] meaning that we want to measure modes two an three.  The 
\textbf{vec} object that corresponds to the projector $\ket{\psi_1}\bra{\psi_1}$ must therefore have kets with 2 modes.
\\
\\
For example, if $\ket{\psi_2}$ is

 $$V2:=\left[ \begin {array}{ccc} 1&&|0000>
\\\noalign{\medskip}x&&|1100>
\\\noalign{\medskip}{x}^{2}&&|2200>
\end {array} \right] $$

and $\ket{\psi_1}$ is

$$V1:=\left[ \begin {array}{ccc} 1&&|10>
\end {array} \right] $$

Then, \\
\frame{ 
\begin{tabular}{cc} \\ 
&    
\begin{minipage}{4in}  
S := \texttt{Project}(V1, [2,3], V2);
\\
\end{minipage}  
\end{tabular}  
}
\\
\\
will return:
$$S = \left[ \begin {array}{ccc} x &&|1100>
\end {array} \right] $$

\item if given (\textbf{matcol 1}, list, \textbf{vec 2}) and say \textbf{matcol 1} and \textbf{vec 2}
describe $M_n$ and $\ket{\psi_2}$ then it will return the \textbf{matcol} object:
$$ \rho = \left(M_{n_{list}} \otimes 1\!\!\:\!{\rm{I}}_{rest}\right)\; \ket{\psi_2}
\bra{\psi_2} \left(M_{n_{list}} \otimes 1\!\!\:\!{\rm{I}}_{rest}\right)^\dagger $$

So for example, taking the same \texttt{V2} as above and the Kraus Operator (or projector):

$$M:= \left[ \begin {array}{ccc} 1&[1,1]&[1,1]\\\noalign{\medskip}1&[0,0]&[0
,0]\end {array} \right] $$

Then, \\
\frame{ 
\begin{tabular}{cc} \\ 
&    
\begin{minipage}{4in}  
S := \texttt{Project}(M, [1,2], V2);
\\
\end{minipage}  
\end{tabular}  
}
\\
\\
will return the state:
$$S =   \left[ \begin {array}{ccc} 1&[0,0,0,0]&[0,0,0,0]\\\noalign{\medskip} \bar{x}  &[0,0,0,0]&[1,1,0,0]
\\\noalign{\medskip}x&[1,1,0,0]&[0,0,0,0]\\\noalign{\medskip}x\bar{ x}  &[1,1,0,0]&[1,1,0,0]\end {array} \right] 
 $$
 which is the density matrix corresponding to:
$$ S =  \left[ \begin {array}{cc} 1&[0,0,0,0]\\\noalign{\medskip}x&[1,1,0,0
]\end {array} \right] $$
\\
\\
\item if given (\textbf{vec 1}, list, \textbf{matcol 2}) and say \textbf{vec 1}, \textbf{matcol 2} 
describe $\ket{\psi_1}$ and $\rho$ respectively, then it will build $\ket{\psi_1}\bra{\psi_1}$ and return the 
\textbf{matcol} object corresponding to:
$$\rho' = \left(\ket{\psi_1}\bra{\psi_1}_{list} \otimes 1\!\!\:\!{\rm{I}}_{rest} \right)\; \rho \;  \left(\ket{\psi_1}\bra{\psi_1}_{list} \otimes 1\!\!\:\!{\rm{I}}_{rest} \right)$$
\\
\\
\item if given (\textbf{matcol 1}, list, \textbf{matcol 2}) and say \textbf{matcol 1} and \textbf{matcol 2}  describe $M_n$ and $\rho$ respectively, then it will return:
$$\rho' =  \left(M_{n_{list}} \otimes 1\!\!\:\!{\rm{I}}_{rest}\right) \; \rho \;  \left(M_{n_{list}} \otimes 1\!\!\:\!{\rm{I}}_{rest}\right)^\dagger$$
\end{enumerate}
(see further down for POVM measurements)\\\\
In a nutshell:
\\
\\
\begin{tabular}{lcl}
\texttt{Project}($\ket{\psi_1}$, list, $\ket{\psi_2}$)& $\rightarrow$& 
$\left(\ket{\psi_1}\bra{\psi_1}_{list} \otimes 1\!\!\:\!{\rm{I}}_{rest}\right) \;\; \ket{\psi_2}$
\\
\texttt{Project}($M_n$, list, $\ket{\psi_2}$) & $\rightarrow$& $  \left(M_{n_{list}} \otimes 1\!\!\:\!{\rm{I}}_{rest}\right)\; \ket{\psi_2}
\bra{\psi_2} \left(M_{n_{list}} \otimes 1\!\!\:\!{\rm{I}}_{rest}\right)^\dagger $
\\
\texttt{Project}($\ket{\psi_1}$, list,$\;\; \rho\;\;$) & $\rightarrow$& $\left(\ket{\psi_1}\bra{\psi_1}_{list} \otimes 1\!\!\:\!{\rm{I}}_{rest} \right) \; \rho \;  \left(\ket{\psi_1}\bra{\psi_1}_{list} \otimes 1\!\!\:\!{\rm{I}}_{rest} \right)$
\\
\texttt{Project}($M_n$, list,$\;\;\rho\;\;$) & $\rightarrow$& $ \left(M_{n_{list}} \otimes 1\!\!\:\!{\rm{I}}_{rest}\right) \; \rho \;  \left(M_{n_{list}} \otimes 1\!\!\:\!{\rm{I}}_{rest}\right)^\dagger$\\\\
\end{tabular}
\\
\\
If we encounter a destructive measurement, it is possible to trace out the measured modes
with the \texttt{Traceout} procedure (cf. appendix).  Also \texttt{POVMresult} traces out
the measured modes (see next section).  We may otherwise calculate the full
trace of a \textbf{mat} or \textbf{matcol} object with \texttt{StateTrace} or multiply \textbf{vec} and \textbf{matcol}
in different orders thanks to \texttt{StateMultiply}.  
\\
\\
\hskip 0.55cm \textbf{The} \texttt{Probability}  \textbf{Procedure}:\\\\
This procedure will calculate the probability of a measurement result or an expected value.
It considers the same cases and objects as the above \texttt{Project} procedure.
It uses the definition,
\\
$$P = \frac{\trace{}{E_n \rho}}{\trace{}{\rho}}$$
\\
and assumes that $\{E_n\}$ safisfy $E_n \geq 0$ and $\sum_n E_n = \Id$ to calculate probabilities.
Therefore, one should verify that these conditions hold for the matrices
describing $E_n$ in order to obtain meaningful probabilities.
Below we show more details for different inputs:
\\
\\
\texttt{Probability}($\ket{\psi_1}$, list, $\ket{\psi_2}$)\hskip 1cm $\rightarrow$ \hskip 0.5cm 
$\displaystyle{ \frac{\trace{}{\left(\ket{\psi_1}\bra{\psi_1}_{list} \otimes 1\!\!\:\!{\rm{I}}_{rest}\right) \;\; \ket{\psi_2}\bra{\psi_2}}}{\trace{}{\ket{\psi_2}\bra{\psi_2}}}}$ 
\\
\\
where ($\ket{\psi_1}$, $\ket{\psi_2}$) are converted to \textbf{matcol} objects in an intermediate
step.  One should pay attention to the choice of the projection operator $\ket{\psi_1}\bra{\psi_1}$.  If it is
not nomalized it can give unphysical values for the probability.
\\
\\
Now for a given POVM or Projector,\\
\texttt{Probability}($E_n$, list, $\ket{\psi_2}$) \hskip 1cm $\rightarrow$ \hskip 0.5cm
$\displaystyle{\frac{\trace{}{(E_{n \; list} \otimes \Id_{rest}) \ket{\psi_2}\bra{\psi_2}}}{\trace{}{\ket{\psi_2}\bra{\psi_2}}} }$ 
\\
\\
\\
Or given a \textbf{vec} and a density operator \textbf{matcol}:\\
\texttt{Probability}($\ket{\psi_1}$, list, $\; \rho\;\;$) \hskip 1cm $\rightarrow$ \hskip 0.5cm
$\displaystyle{\frac{\trace{}{\left(\ket{\psi_1}\bra{\psi_1}_{list} \otimes 1\!\!\:\!{\rm{I}}_{rest}\right) \rho}}{\trace{}{\rho}} }$ 
\\
\\
where $\ket{\psi_1}$ has been converted to a \textbf{matcol} object.\\
\\
\\
And finally for a POVM and a density operator:\\
\\
\\
\texttt{Probability}($E_n$, list, $\;\rho\;\;$) \hskip 1cm $\rightarrow$\hskip 0.5cm 
$\displaystyle{\frac{\mathrm{Tr}\left\{
(E_{n\;list} \otimes 1\!\!\:\!{\rm{I}}_{rest}) \;\rho\;
\right\}}
{\texttt{Tr}\left\{\; \rho\;\right\}}}$
\\
\\
\\
\subsubsection{POVM measurements}
Quantavo posesses a procedure to describe POVM measurements.  If our state before the measurement
is $\ket{\psi}$ or $\rho$, and the POVM elements are described by the set $\{E_m\}$, satifying
$\sum_m E_m =1\!\!\:\!{\rm{I}}$ and $E_m \geq 0$.  Then the state after the measurement will be:
$$\rho' = \trace{i,j,..k}{E_m \; \rho}$$
or
$$\rho' = \trace{i,j,..k}{E_m \ket{\psi}\bra{\psi}}$$
assuming $\trace{}{\rho} = 1$ or $\trace{}{\ket{\psi}\bra{\psi}} = 1$.  
\\
\\
This will be implemented by the procedure \texttt{POVMresult}
which will take as inputs,
\\
\\
\texttt{POVMresult}(\textbf{matcol}, $List$, \textbf{matcol}/\textbf{vec})
\\
\\
and implement the operation,
\\
\\
\texttt{POVMresult}($E_m$, $List$, $\rho$)
= $\displaystyle{\frac{\trace{List}{\left(E_{m List} \otimes \Id_{rest}\right) \; \rho }}{\trace{}{\rho}}}$\\
where `$rest$' are all the indexes not included in `$List$'.  
\\
\\
\subsubsection{Declaring POVMs}
The \textbf{matcol} objects that represent POVM operators need to be declared.
One option is to declare them as standard states and convert them to \textbf{matcol}.
Since avalanche photo diode detectors (APDs) are a standard tool in quantum optics,
an interactive tool to declare them is also provided.  Executing,
\\
\\
\frame{
\begin{tabular}{cc} \\
&
\begin{minipage}{4.5in}
APD();
\\
\end{minipage}
\end{tabular}
}
\\\\
Will bring up an interactive menu, where we can choose if we want the
$\ket{0}\bra{0}$ or $\displaystyle{\Id - \ket{0}\bra{0}}$ event, how many photons we
will consider and what loss should be added to it.  For example choosing the input \textbf{(0, r, 4)} will return,
$$\pi_0 =\left[ \begin {array}{ccc} 1&[0]&[0]\\\noalign{\medskip}{r}^{2}&[1]&[
1]\\\noalign{\medskip}{r}^{4}&[2]&[2]\\\noalign{\medskip}{r}^{6}&[3]&[
3]\\\noalign{\medskip}{r}^{8}&[4]&[4]\end {array} \right] 
$$
and choosing \textbf{(1, r, 4)} will return,
$$\pi_1 = \left[ \begin {array}{ccc} 1-{r}^{2}&[1]&[1]\\\noalign{\medskip}1-{r}
^{4}&[2]&[2]\\\noalign{\medskip}1-{r}^{6}&[3]&[3]\\\noalign{\medskip}1
-{r}^{8}&[4]&[4]\end {array} \right]$$
recovering the expected $\pi_0 + \pi_1 = \Id$.  Note the convention
for the BS in front of the detector for which $r=0$ is a perfect detector and $r^2+t^2=1$.
\\
\\
These are therefore the main tools to describe and simulate quantum measurements in this framework.

\subsubsection{State Properties}

Some questions that will interest us will concern hermiticity, normalization and
entanglement measures.  So far, {\sf Quantavo} has the following usefull procedures:
\\
\\
\texttt{IsHermitian} to check for selfadjointness, 
\\
\texttt{IsNormalized} to check for normalization,
\\
\texttt{StateNormalize} to normalize \textbf{mat}, \textbf{matcol} or \textbf{vec}
\\
\texttt{StatePartialTranspose} to partial transpose $\rho^{\Gamma}$
\\
\texttt{Negativity} to calculate the negativity as the sum of all negative eigenvalues 
of the partial transposed density matrix.
\\
\texttt{LogNegativity} to calculate the Logarithmic Negativity.
\\
\texttt{StateApprox} to do symbolic or numeric approximations transforming our \textbf{matcol} or \textbf{vec} state.
\\
\\

For practical examples on how to use them one can refer to the following section.
Otherwise, a detailed dictionary of procedures can be found in the appendix.

\section{Practical Example}\label{example}
\subsection{Squeezed state photon subtraction}

\begin{figure}[ht]
\centerline{\includegraphics[width=6cm]{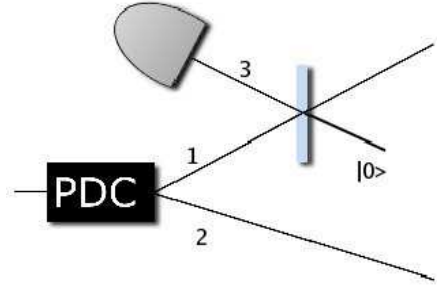}}
\caption{Setup of the photon subtraction.\label{subtraction}  }
\end{figure}

Our initial state is a pure two mode squeezed state that 
can be described by 
$|\psi_\lambda\rangle = \sqrt{1-\lambda^2}\sum_{n=0}^{\infty} \lambda^n |n,n\rangle$.
We would like to declare it and propagate it through the circuit presented in fig. \ref{subtraction}.
We observe that our initial state has three modes, one of which is a vacuum mode.
We can work out as an example the state with up to 4 photons:
\\
\\
\frame{
\begin{tabular}{cc} \\
&
\begin{minipage}{4.5in}

\# Create a truncated two mode squeezed vacuum state\\
V := SqueezedVac(2, 5, lambda);\\ 

\# Then add the vacuum mode:\\
V := TensorVac(V,1);\\
\end{minipage}
\end{tabular}
}
\\\\
This will output state:
$$V :=   \left[ \begin {array}{cc} 1&[0,0,0]\\\noalign{\medskip}\lambda&[1,1,0
]\\\noalign{\medskip}{\lambda}^{2}&[2,2,0]\\\noalign{\medskip}{\lambda
}^{3}&[3,3,0]\\\noalign{\medskip}{\lambda}^{4}&[4,4,0]\end {array}
 \right] 
 $$

We then apply the corresponding Beamsplitter transformation (if ``$d$'' changes from
the BS transformation, it is automatically recalculated after the BS operation and reset
to its new value.  In this case it doesn't change).
\\
\\
\frame{
\begin{tabular}{cc} \\

&
   
\begin{minipage}{4.5in}

V1:=BS(V,1,3);\\

\end{minipage}
\end{tabular}
}
\\\\
which returns, if displayed with ``Display State'', that is \texttt{Dstate}(V1), the following output:

$$V1= \left[ \begin {array}{ccc} 1& &``|000>''
\\\noalign{\medskip}\lambda\,t& &``|110>''
\\\noalign{\medskip}{\lambda}^{2}{t}^{2}&&
``|220>''\\\noalign{\medskip}{\lambda}^{3}{t}^{3}&
&``|330>''\\\noalign{\medskip}{\lambda}^{4}{t}^{4}&
&``|440>''\\\noalign{\medskip}\lambda\,r&
&``|011>''\\\noalign{\medskip}\sqrt {2}{\lambda}^{2}tr&
&``|121>''\\\noalign{\medskip}\sqrt {3}{\lambda
}^{3}{t}^{2}r&&``|231>''\\\noalign{\medskip}2\,
{\lambda}^{4}{t}^{3}r&&``|341>''
\\\noalign{\medskip}{\lambda}^{2}{r}^{2}&&
``|022>''\\\noalign{\medskip}\sqrt {3}{\lambda}^{3}t{r}^{2}&
&``|132>''\\\noalign{\medskip}\sqrt {6}{\lambda}^{4}{t}
^{2}{r}^{2}&&``|242>''\\\noalign{\medskip}{
\lambda}^{3}{r}^{3}&&``|033>''
\\\noalign{\medskip}2\,{\lambda}^{4}t{r}^{3}&&
``|143>''\\\noalign{\medskip}{\lambda}^{4}{r}^{4}&
&``|044>''\end {array} \right] 
$$
\\
\textbf{Measurements}:
\\
\\
First we will consider the case of a perfect photon number resolving detector.
The projector $\ket{1}\bra{1}$ can simply be introduced as:
\\
\\
\frame{
\begin{tabular}{cc} \\

&
   
\begin{minipage}{4.5in}

Proj:=Matrix([1,[1]]);\\

\end{minipage}
\end{tabular}
}
\\
\\
And the state after the measurement will be obtained with:
\\
\\
\frame{
\begin{tabular}{cc} \\

&
   
\begin{minipage}{4.5in}

V2:=\texttt{Project}(Proj,[3],V1);\\

\end{minipage}
\end{tabular}
}
\\
\\
returning:
$$
 V2 = \left[ \begin {array}{cc} \lambda\,r&[0,1,1]\\\noalign{\medskip}
\sqrt {2}{\lambda}^{2}tr&[1,2,1]\\\noalign{\medskip}\sqrt {3}{\lambda}
^{3}{t}^{2}r&[2,3,1]\\\noalign{\medskip}2\,{\lambda}^{4}{t}^{3}r&[3,4,
1]\end {array} \right] 
$$
We may now trace-out the measured mode. Tracing out can 
then be done as follows:
\\
\\
\frame{
\begin{tabular}{cc} \\

&
   
\begin{minipage}{4.5in}
M1:=\texttt{Traceout}(V2,3);\\

\end{minipage}
\end{tabular}
}
\\
\\

We may also simulate a measurement with an avalanche photo-diode described by the Kraus Operator:
$\hat{O} = 1\!\!\:\!{\rm{I}} -\ket{0}\bra{0}$.  That is, if the state ``V1'' represents the state $\rho$, we want to find the state $\rho'$ resulting from the measurement:

$$\rho' \sim \left(\hat{O}_3  \otimes 1\!\!\:\!{\rm{I}}_{1,2}\right) \rho \left(\hat{O}_3 \otimes  1\!\!\:\!{\rm{I}}_{1,2}\right)^\dagger$$

For this simple example we have at most 4 photons so we will approximate $\hat{O} \simeq \ket{1}\bra{1}+\ket{2}\bra{2}+
\ket{3}\bra{3}+\ket{4}\bra{4}$.
Therefore to measure mode $3$ we construct the associated POVM which can then be expressed as a \textbf{matcol} object:
\\
\\
\frame{
\begin{tabular}{cc} \\
&
\begin{minipage}{4.5in}
POVM := Matrix(4, 3);\\\\
for i from 1 to 4 do\\
POVM[i,1]:=1:\\
POVM[i,2]:=[i]:\\
POVM[i,3]:=[i]:\\
od:\\
\end{minipage}
\end{tabular}
}
\\
\\
And the state after a `click' in the detector (tracing out this inaccessible mode) will be:
\\
\\
\frame{
\begin{tabular}{cc} \\

&
   
\begin{minipage}{4.5in}

M2:=Project(POVM,[3],V1):\\
M3:=Traceout(M2,3):\\

\end{minipage}
\end{tabular}
}
\\
\\
or directly
\\
\\
\frame{
\begin{tabular}{cc} \\

&
   
\begin{minipage}{4.5in}

M3:=POVMresult(POVM,[3],V1):\\
\end{minipage}
\end{tabular}
}
\\
\\
\textbf{Approximations}:
\\
\\
M3 has 30 coefficients.  We may want to use the approximation procedures to simplify 
our calculations.  If for example we know that $r<<1$ and $\lambda << 1$ we may wish
to delete all terms in the density matrix for which the coefficients containing $\mathrm{r}^n \lambda^m$ satisfy $n+m > 7$
(therefore getting rid of all the small terms up to the chosen order).  The procedure \texttt{StateApprox} 
can be used for this purpose.  This way, 
\\
\\
\frame{
\begin{tabular}{cc} \\
&
\begin{minipage}{4.5in}
M4:=RealDomain:-simplify(M3): \# assume real variables for simplicity\\
M5:=\texttt{StateApprox}(M4,[lambda,r], 7);\\
\end{minipage}
\end{tabular}
}
\\
\\
Will deliver
$$
M5 = \left[ \begin {array}{ccc} {\lambda}^{2}{r}^{2}&[0,1]&[0,1]
\\\noalign{\medskip}\sqrt {2}{\lambda}^{3}t{r}^{2}&[0,1]&[1,2]
\\\noalign{\medskip}{\lambda}^{4}{t}^{2}{r}^{2}\sqrt {3}&[0,1]&[2,3]
\\\noalign{\medskip}2\,{\lambda}^{5}{t}^{3}{r}^{2}&[0,1]&[3,4]
\\\noalign{\medskip}\sqrt {2}{\lambda}^{3}t{r}^{2}&[1,2]&[0,1]
\\\noalign{\medskip}2\,{\lambda}^{4}{t}^{2}{r}^{2}&[1,2]&[1,2]
\\\noalign{\medskip}\sqrt {6}{\lambda}^{5}{t}^{3}{r}^{2}&[1,2]&[2,3]
\\\noalign{\medskip}{\lambda}^{4}{t}^{2}{r}^{2}\sqrt {3}&[2,3]&[0,1]
\\\noalign{\medskip}\sqrt {6}{\lambda}^{5}{t}^{3}{r}^{2}&[2,3]&[1,2]
\\\noalign{\medskip}2\,{\lambda}^{5}{t}^{3}{r}^{2}&[3,4]&[0,1]
\end {array} \right] 
$$
\\
See the dictionary in the appendix for details on how to use the \texttt{StateApprox} procedure.
\\
\\
\textbf{Entanglement}\\\\
We may now compute different properties of this state as for example the Negativity.  We take 
the state after the detection of exactly one photon and obtain:
\\
\\
\frame{
\begin{tabular}{cc} \\
&
\begin{minipage}{4.5in}
Neg:=simplify(Negativity(M1)) assuming t::positive, r::positive, lambda::positive;\\
\end{minipage}
\end{tabular}
}
\\
which will return:
\begin{eqnarray*}
Neg &=& {\frac {\lambda\,t \left( 2\,{\lambda}^{2}{t}
^{2}+\sqrt {2}+\sqrt {3}\lambda\,t+\sqrt {3}{\lambda}^{2}{t}^{2}\sqrt 
{2}+2\,{\lambda}^{3}{t}^{3}\sqrt {2}+2\,{\lambda}^{4}{t}^{4}\sqrt {3}
 \right) }{4\,{\lambda}^{6}{t}^{6}+3\,{\lambda}^{4}{t}^{4}+2\,{\lambda
}^{2}{t}^{2}+1}}
\end{eqnarray*}
The eigenvalues can be hard to find if our state has symbolic
entries with complex conjugation.  A previous simplification (assuming 
real variables) can help, but other solutions are possible.
This explicit expression allows us to plot different values of the negativity for
various ranges of parameters: \\\\
\frame{
\begin{tabular}{cc} \\
&
\begin{minipage}{4.5in}
plot3d(Neg($\lambda$,t,$\sqrt{1-t^2}$), $\lambda$=0..1,t=0..1, axes=boxed);\\

\end{minipage}
\end{tabular}
}
\\
will display the plot in fig \ref{negativity}.:

\begin{figure}[ht]
\centerline{\includegraphics[width=6cm]{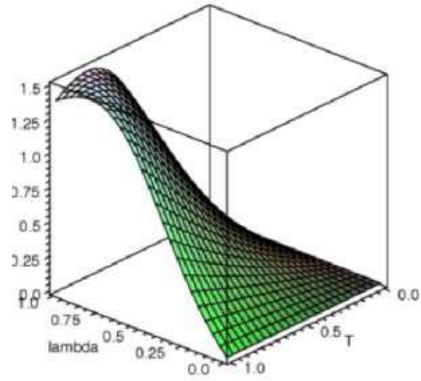}}
\caption{Logarithmic Negativity (t, $\lambda$).\label{negativity}  }
\end{figure}

Further details, examples and documentation can be found on the website of the
Imperial College London quantum information group,\\
\url{http://www.imperial.ac.uk/quantuminformation/research/downloads}

\section*{Acknowledgements}
This project was supported by EPSRC grant EP/C546237/1 and by the Integrated Project Qubit Applications
(QAP) supported by the IST directorate as Contract Number 015848.  The author would also like to thank 
Alain Le Stang and the members of \texttt{Mapleprimes.com} for useful suggestions.
\newpage
\appendix

\section{Dictionary of procedures:}

A quick reference list is provided and an alphabetically ordered
dictionary follows.
\\
\\
\begin{tabular}{|c|c|c|}
    \hline
Objects &  Appearance & Declaration       \\
    \hline
    &&\\
\textbf{vec}
&
$\psi = \left[ \begin {array}{cc}
1&[0,0,0]\\\noalign{\medskip}\lambda&[1,1,0
]\\\noalign{\medskip}{\lambda}^{2}&[2,2,0]\\\noalign{\medskip}{\lambda
}^{3}&[3,3,0]\end {array} \right] $
&
\begin{tabular}{l}
vec:=\texttt{CoherentState}(K,d,$\alpha$)\\
vec:=\texttt{SqueezedVac}(K,d,$\lambda$)\\
vac:=\texttt{Vac}(K);\\
vec1:=\texttt{TensorVac}(vec,m)\\
vec1:=\texttt{TensorProduct}(V,[1,2],W,[3,4])\\
vec:=Trim(V)\\
\end{tabular}
\\
&&
\\
\hline &&
\\
\textbf{mat}   & $\rho = \left[ \begin {array}{ccc}
0&[0,0,0]&[1,1,0]
\\\noalign{\medskip}[0,0,0]&1&\bar{\lambda}
\\\noalign{\medskip}[1,1,0]&\lambda&\lambda\,\bar{\lambda}
\end {array}
\right] $
&
\begin{tabular}{l}
mat:=\texttt{vec2mat}(vec)\\
mat:=\texttt{matcol2mat}(matcol)\\
or direct declaration of the\\
Matrix\\
\end{tabular}
\\
&&
\\
\hline
&&\\
\textbf{matcol}   &
$\rho_{matcol} = \left[ \begin {array}{ccc}
1&[0,0,0]&[0,0,0]
\\\noalign{\medskip} \bar{\lambda}&[0,0,0]&[1,1,0]
 \\\noalign{\medskip} \lambda&[1,1,0]&[0,0,0]
 \\\noalign{\medskip} \lambda\,\bar{\lambda} &[1,1,0]&[1,1,0]
 \\\noalign{\medskip}{\lambda}^{2}  \bar{\lambda}^{2}&[2,2,0]&[2,2,0]
\end {array}
\right] $ &
\begin{tabular}{l}
matcol:=\texttt{vec2matcol}(vec)\\
matcol:=\texttt{mat2matcol}(mat)\\
or direct declaration of the\\
3 column Matrix\\
\end{tabular}
\\
&&
\\
\hline
\end{tabular}
\\
\\
\\
\\
ready made states are: $\begin{array}{lcc}
\texttt{CoherentState}(K,d,\alpha)&\sim &\displaystyle{
\sum_{n=0}^{d-1} \frac{\alpha^n}{\sqrt{n!}} \ket{n}^{\otimes K}}
\\\texttt{SqueezedVac}(K,d,\lambda)& \sim& \displaystyle{\sum_{n=0}^{d-1}
\lambda^n \ket{n}^{\otimes K}}
\\\texttt{TensorVac}(\textbf{vec},m)&:& \displaystyle{\ket{\phi} \rightarrow
\ket{\phi}\otimes \ket{0}^{\otimes m}}\\
\\\texttt{IdentityState(d,K)} & \sim & \displaystyle{\Id_{d^K \times d^K} }\\
\end{array}$
\section*{Common Procedures:}
\vskip 1cm
\hskip -1.8cm
\begin{tabular}{clc}
    \hline
   &\vline&\\
\verb"Linear Optics":  &\vline&  \verb"Measurements" \\
    &\vline&
    \\
\hline
&\vline&\\
  $\begin{array}{l}
\texttt{BS}(\mathbf{vec}/\mathbf{matcol},\mathrm{i},\mathrm{j})\\
\texttt{myBS}(\mathbf{vec}/\mathbf{matcol},\mathrm{i},\mathrm{j},\mathrm{t},\mathrm{r})\\\\
\texttt{PS}(\mathbf{vec}/\mathbf{matcol},\mathrm{i},\mathrm{\phi})\\\\
\texttt{BuildUnitary}([List\; of Lists])\\ 
\texttt{UnitaryEvolution}(U,\mathbf{vec}/\mathbf{matcol})\\
\end{array}$
&\vline&
 $\begin{array}{l}
\texttt{Project}(\mathbf{vec}/\mathbf{matcol}, [List], \mathbf{vec}/\mathbf{matcol}) \\
\texttt{Probability}(\mathbf{vec}/\mathbf{matcol}, [List] \mathbf{vec}/\mathbf{matcol})\\
\texttt{Traceout}(\mathbf{mat}/\mathbf{matcol}, \mathrm{i})\\
\texttt{POVMresult}(\mathbf{matcol},[List], \mathbf{vec}/\mathbf{matcol})\\
\texttt{APD}\left(\{0,1\}, r, d+1 \right)\\
\end{array}$\\
&\vline&\\
\hline
\\
\\
    \hline
    &\vline&\\
\verb"Display"  &\vline& \verb"Algebraic Operations" \\
&\vline&\\
\hline
&\vline&\\
 $\begin{array}{l}
\texttt{Dstate}(\mathbf{vec}/\mathbf{matcol}/\mathbf{mat})\\ \\
\texttt{PlotState}(\mathbf{vec}/\mathbf{matcol}/\mathbf{mat},\mathrm{width,height})\\

\end{array}$
&\vline& $\begin{array}{l}
\texttt{StateApprox}(\mathbf{vec}/\mathbf{matcol}, list,N)\\
\texttt{StateMultiply}(\mathbf{vec}/\mathbf{matcol},\mathbf{vec}/\mathbf{matcol})\\
\texttt{StateComplexConjugate}(\mathbf{mat}/\mathbf{matcol})\\
\texttt{StateNorm}(\mathbf{vec})\\
\texttt{StateNormalize}(\mathbf{vec}/\mathbf{mat}/\mathbf{matcol})\\
\texttt{StateTrace}(\mathbf{mat}/\mathbf{matcol})\\
\texttt{Traceout}(\mathbf{mat}/\mathbf{matcol},\mathrm{i})\\
\texttt{TensorProduct}(\mathbf{vec}/\mathbf{matcol},list,\mathbf{vec}/\mathbf{matcol},list)\\
\texttt{StatePartialTranspose}(\mathbf{mat}/\mathbf{matcol},\mathrm{i})\\
\texttt{deltaK}(\mathrm{i},\mathrm{j})\\
\end{array}$\\
&\vline&\\
\hline

\\
\\
    \hline
    &\vline&\\
\verb"State Properties:" &\vline& \verb"Entanglement & Energy" \\
&\vline&\\
\hline
&\vline&\\
 $\begin{array}{l}
\texttt{IsHermitian}(\mathbf{matcol}/\mathbf{mat})\\
\texttt{IsNormalized}(\mathbf{vec}/\mathbf{matcol}/\mathbf{mat})\\
\texttt{StateNorm}(\mathbf{vec})\\
\texttt{StateSort}(\mathbf{vec}/\mathbf{matcol})\\
\texttt{FindKnd}(\mathbf{vec}/\mathbf{matcol}/\mathbf{mat})\\
\end{array}$
&\vline& $\begin{array}{l}
\texttt{Negativity}(\textbf{vec}/\mathbf{matcol}/\mathbf{mat})\\
\texttt{LogNegativity}(\textbf{vec}\mathbf{matcol}/\mathbf{mat})\\
\texttt{Entropy}(\mathbf{vec}/\mathbf{matcol})\\
\texttt{Energy}(\mathbf{vec}/\mathbf{matcol})\\
\end{array}$\\
&\vline&\\
\hline
\end{tabular}
\\
\\
\\
\\
\newpage
\begin{LARGE}A\end{LARGE}\\
\\
\defi{APD}{0/1,r,N}{This procedure is interactive and is called by executing,
\\
\\
APD();
\\
\\
It declares the POVM describing a lossy avalanche photo diode 
detector (APD).
The input ``0'' or ``1'' will select between the \textsl{no-click} or \textsl{click} events
respectively.  ``$r$'' will be the amplitude $r$ of the reflectivity $R=r^2$ of the BS in front of the detector characterizing its loss, and $N$ will be the maximum number of photons.
For example choosing the input \textbf{(0, r, 4)} will return,
$$\pi_0 =\left[ \begin {array}{ccc} 1&[0]&[0]\\\noalign{\medskip}{r}^{2}&[1]&[
1]\\\noalign{\medskip}{r}^{4}&[2]&[2]\\\noalign{\medskip}{r}^{6}&[3]&[
3]\\\noalign{\medskip}{r}^{8}&[4]&[4]\end {array} \right] 
$$
and choosing \textbf{(1, r, 4)} will return,
$$\pi_1 = \left[ \begin {array}{ccc} 1-{r}^{2}&[1]&[1]\\\noalign{\medskip}1-{r}
^{4}&[2]&[2]\\\noalign{\medskip}1-{r}^{6}&[3]&[3]\\\noalign{\medskip}1
-{r}^{8}&[4]&[4]\end {array} \right]$$
recovering the expected $\pi_0 + \pi_1 = \Id$.
\\
\\}
{($\{0,1\}$, r $\in [0,1]$, positive integer)}{Matrix}{Quantavo, LinearAlgebra}
\\
\\
\begin{LARGE}B\end{LARGE}\\
\\
\defi{BS}{vec/matcol, $i$, $j$}{This will effectuate the Beam Splitter transformation:

\begin{eqnarray}
\left(
\begin{array}{c}
 a_i'\\ a_j'
\end{array} \right) 
=
\left( 
\begin{array}{cc}
t & r\\
-r & t
\end{array}
\right)
\left(
\begin{array}{c}
 a_i\\ a_j
\end{array} \right) \\
\left(
\begin{array}{c}
 b_i'\\ b_j'
\end{array} \right) 
=
\left( 
\begin{array}{cc}
t^* & r^*\\
-r^* & t^*
\end{array}
\right)
\left(
\begin{array}{c}
 b_i\\ b_j
\end{array} \right). 
\end{eqnarray}
\\
for the chosen modes.  That is eq. (8) for objects of type \textbf{vec} and eq. (8) and (9) for objects
of type \textbf{matcol}.  The 
value of `$d$' and `$K$' are evaluated and reset to the actual value after this operation.
}{(Matrix, integer, integer)}{Matrix}{Quantavo, LinearAlgebra, poly2matcol, matcol2poly, matcolBS, vecBS}
\\
\\
\defi{BuildUnitary}{list of lists}{This procedure will build a $K \times K$ unitary matrix corresponding 
to a linear optics circuit consisting of beam splitters (BS) and phase shifters (PS).  
To do so a list of lists must be provided.  The lists
inside the list can have 2, 3, or 4 elements, and should be in the same order as we want to apply those 
transformations in the circuit.  Lists with 2 elements will be considered PS and lists
with 3 or 4 elements as BS:\\
\\
The list [i, $\xi$] will implement a PS on mode $i$.
\\
The list [i,j,t] will implement a BS on modes i and j with transmittivity $t^2=T$ and reflectivity $R=1-T=1-t^2$.
\\
The list [i,j,t,r] will implement a BS on modes i and j with transmittivity $t^2=T$ and reflectivity $r^2=R$.
\\
\\
To implement the transformations one after another we can give for example the input\\list = [[$i_1$, $\xi$],[$i_2$, $j_2$, t],[$i_3$, $j_3$, t, r],[$i_3$,$\phi$]].}{List}{Matrix}{Quantavo, LinearAlgebra}
\\
\\
\begin{LARGE}C\end{LARGE}\\
\\
\defi{CoherentState}{m,d,$\alpha$}{This procedure builds an object of type \textbf{vec} that describes the state $\ket{\phi} = \sum_{n=0}^{d-1} \frac{\alpha^n}{\sqrt{n!}} \ket{n}^{\otimes m}$.  It is not normalized}{(whole number, whole number)}{Matrix}{Quantavo, LinearAlgebra}
\\
\\
\begin{LARGE}D\end{LARGE}\\
\\
\defi{DP}{Matrix, Matrix}{This procedure makes the Kronecker/Direct/Tensor product between any two matrices.
As long as it is declared as a matrix, it can also make the Kronecker product between vectors and matrices.  The vector, however would have to be declared as a Matrix. For example as V := Matrix([[1], [2], [0], [a]]).}{Matrix}{Matrix}{Quantavo, LinearAlgebra}
\\
\\
\defi{DeltaK}{i, j}{This procedure takes the Kronecker Delta from two inputs.
It works both symbolically and numerically.  For example:\\
deltaK(1,1)=1, deltaK(1,2)=0,\\
deltaK($a$,$a$)=1,  (even if $a$ is not specified).\\
deltaK(a,b); will return Quantavo:-deltaK(a, b)\\
However, if we execute ``a:=b;'' and evaluate it again it will be $1$}{(string or number,  
string or number)}{0,1, unevaluated string}{Quantavo}
\\
\\
\defi{Dbra}{List}{This is a Display procedure.  When given 
a list which stands for a bra, for example List =[0,1,3] it builds a bra for display and returns
$\langle 013|$.}{List}{$\langle$ string$|$}{Quantavo}.
\\
\\
\defi{Dbraket}{List, List}{This is a Display procedure.  When given 
two lists which stand for a ket and a bra it returns a $|ket\rangle\langle bra|$. for example IList=[0,1,1] and JList =[0,1,3] then
\texttt{Dbraket}(IList,JList) =$|011\rangle \langle 013|$.}{(List, List)}{$|$string$\rangle \langle$ string$|$}{Quantavo}
\\
\\
\defi{Dket}{List}{This is a Display procedure:  When given 
a list which stands for a ket, for example List =[0,1,3] it builds a ket for display and returns
$|013\rangle$.}{List}{$|$ string$\rangle$}{Quantavo}
\\
\\
\defi{Dstate}{vec/mat/matcol}{ This is a Display procedure.  When given
either a \textbf{mat}, \textbf{vec} or \textbf{matcol} object it transforms 
the lists of modes into \textit{bras} and \textit{kets}.  For example if it is 
given the object \textbf{mat}:
$$ \rho=\left[ \begin {array}{ccc} 0&[0,0,0,0]&[1,1,0,0]\\\noalign{\medskip}[0
,0,0,0]&1&\bar{\lambda}  \\\noalign{\medskip}[
1,1,0,0]&\lambda&\lambda\,\bar{\lambda}
\end {array} \right] $$
it will display bras and kets in the following way:
$$\mathrm{{\tt Dstate}}(\rho) = \left[ \begin {array}{ccc} 0&<0000|&<1100|
\\\noalign{\medskip}|0000>&1&\bar{ \lambda}
 \\\noalign{\medskip}|1100>&\lambda&\lambda\,\bar{ \lambda} \end {array} \right] $$}{Matrix}{Matrix}{Quantavo, LinearAlgebra}
\\
\\
\begin{LARGE}E\end{LARGE}
\\
\\
\defi{Energy}{vec/matcol}{This will output the expected value of the energy of our state defined as,
$$\langle \hat{E} \rangle = \bra{\psi} \hat{E} \ket{\psi}$$
with
$$\langle \hat{E} \rangle = \hbar \nu \trace{}{(a^\dagger a + 1/2) \rho}$$
or for multipartite states,
$$\langle \hat{E} \rangle = \hbar \nu \trace{}{(\hat{N}_1+\hat{N}_2+...+\hat{N}_K + K/2) \rho}$$
}{Matrix}{Number or Analytic Expression}{Quantavo, LinearAlgebra}
\\
\\
\defi{Entropy}{vec/matcol}{This will output the entropy of our state defined as,
$$\mathcal{S} = \sum_i \lambda_i log_2 (\lambda_i)$$
where $\lambda_i$ are the eigenvalues.  The procedure transforms the
objects \textbf{vec} or \textbf{matcol} into a \textbf{mat} density matrix
and then finds the eigenvalues.
}{Matrix}{Number or Analytic Expression}{Quantavo, LinearAlgebra}
\\
\\
\defi{EvalState}{vec/matcol}{This simply applies the Maple function
\texttt{evalf} to the first column of our \textbf{vec} or \textbf{matcol} objects.
}{Matrix}{Number or Analytic Expression}{Quantavo, LinearAlgebra}
\\
\\
\begin{LARGE}F\end{LARGE}
\\
\\
\defi{findKnd}{vec/matcol/mat}{This will search through the given state to find the number of modes $K$ and 
the dimension of the modes $d$.  It updates these global variables
with the $K$ and $d$ found and displays them as output}{Matrix}{(K,d)}{Quantavo, LinearAlgebra} 
findKnd
\\
\\
\begin{LARGE}I\end{LARGE}\\
\\
\defi{IdentityState}{Nr of photons, Nr of modes}{This will declare an identity
matrix represented as a \textbf{matcol} object.  For example, for (2,2) as input
we will obtain $$
\Id = \left[ \begin {array}{ccc} 1&[0,0]&[0,0]\\\noalign{\medskip}1&[0,1]&[0
,1]\\\noalign{\medskip}1&[0,2]&[0,2]\\\noalign{\medskip}1&[1,0]&[1,0]
\\\noalign{\medskip}1&[1,1]&[1,1]\\\noalign{\medskip}1&[1,2]&[1,2]
\\\noalign{\medskip}1&[2,0]&[2,0]\\\noalign{\medskip}1&[2,1]&[2,1]
\\\noalign{\medskip}1&[2,2]&[2,2]\end {array} \right]
$$.}{(integer,integer)}{Matrix}{Quantavo, LinearAlgebra}
\\ 
\\
\defi{indexstate}{vec/matcol}{This procedure will transform the last two columns of a \textbf{vec} or \textbf{matcol} object.  Each list containing the number of photons in each mode will be translated into a natural number with the procedure \texttt{VectorRow}(List,d).  This number will indicate the order in which the modes are ordered.  In effect it does a number basis change from a $d$-base to a 10-base.  
For example,\\
$\left[ \begin {array}{cc} 1&[0,0]\\\noalign{\medskip}\xi&[1,1]
\\\noalign{\medskip}{\xi}^{2}&[2,2]\end {array} \right] $
$\longrightarrow$
$\displaystyle{\left[ \begin {array}{cc} 1&1\\\noalign{\medskip}\xi&5
\\\noalign{\medskip}{\xi}^{2}&9\end {array} \right] }$
}{3 column Matrix}{3 column Matrix}{Quantavo, LinearAlgebra}
\\
\\
\defi{IsHermitian}{matcol/mat}{
This procedure verifies if its input describes a Hermitian state.
Therefore if the density matrix it represents verifies $\rho^+ = \rho$.  If
it is Hermitian the value returned will be $0$, otherwise it will be $1$.  Maple might
not recognize products of conjugated complex variables as equal, so one has to make
sure they are simplified.}{Matrix}{0 or 1 and printed answer}{Quantavo, LinearAlgebra}
\\
\\
\defi{IsNormalized}{vec/matcol/mat}{This procedure checks if the considered object is normalized.  Therefore for \textbf{vec} if $\langle \psi | \psi \rangle = 1 $, for \textbf{matcol} and \textbf{mat} if Tr$(\rho)$=1.  If the state is not normalized it will return its norm or trace.}{Matrix}{printed answer}{Quantavo, LinearAlgebra}
\\
\\
\begin{LARGE}L\end{LARGE}\\
\\
\defi{LogNegativity}{vec/matcol/mat}{This procedure evaluates the Logarithmic negativity according to: 
$$LogNegativity(M) := log_2 (2*Negativity(M)+1)$$
See \texttt{Negativity} for further details.}{Matrix}{Expression or number}{Quantavo, Negativity}
\\
\\
\begin{LARGE}M\end{LARGE}\\
\\
\defi{matcol2mat}{matcol}{This translation procedure takes an object of type \textbf{matcol} and converts it into an object of type \textbf{mat}.  To do so, it sorts the \textbf{matcol} object, counts the number of different bras and kets needed and constructs the hermitian matrix \textbf{mat} with the least number of zeros containing all the elements of
\textbf{matcol} }{Matrix}{Matrix}{Quantavo, LinearAlgebra, StateSort}
\\
\\
\defi{matcol2poly}{matcol}{This translation procedure takes an object of type \textbf{matcol} and converts it into an object of type \textbf{poly}.  To do so it
looks at the $\ket{\mathrm{ket}}$ and  $\bra{\mathrm{bra}}$ of each row of the Matrix and builds the associated monomial with its corresponding coefficient. }{Matrix}{Polynomial}{Quantavo, LinearAlgebra}
\\
\\
\defi{mat2matcol}{mat}{This translation procedure takes an object of type \textbf{mat} and converts it into an object of type \textbf{matcol}}{Matrix}{Matrix}{Quantavo, LinearAlgebra, Trim}
\\
\\
\defi{mat2poly}{mat}{This translation procedure takes an object of type \textbf{mat} and converts it into an object of type \textbf{poly}}{Matrix}{polynomial}{Quantavo, LinearAlgebra}
\\
\\
\defi{modesmatcol}{matcol with numbers}{This procedure is the inverse of the procedure \texttt{indexmatcol}.  It will transform a 3 column matrix, that has numbers in the 2nd and 3rd column to one that has the equivalent modes (lists with the number of photons)}{3 column Matrix}{3 column Matrix}{Quantavo, LinearAlgebra}
\\
\\
\defi{myBS}{matcol/vec, $i$, $j$, t, r}{This procedure is the same as \texttt{BS}, but it gives the user the option to choose the transmittivity $T=t^2$ and reflectivity $R=r^2$ of the beam splitter.
}{Matrix}{Matrix}{Quantavo, LinearAlgebra, poly2matcol, matcol2poly, mymatcolBS, myvecBS}
\\
\\  
\begin{LARGE}N\end{LARGE}\\
\\
\defi{Negativity}{vec/matcol/mat}{This procedure finds the eigenvalues of the partial transposed density matrix,  and calculates the sum of all the corresponding non-negative eigenvalues (it also divides by the Trace of the state in case it is not normalized):
$$Negativity =\frac{1}{Tr(\rho)} \sum_{i} \frac{|\lambda_i|-\lambda_i}{2} $$ 
where $\lambda_i$ are the eigenvalues of the partially transposed density matrix}
{Matrix}
{Expression or number}
{Quantavo, LinearAlgebra, StatePartialTranspose}
\\
\\
\begin{LARGE}P\end{LARGE}\\
\\
\\
\defi{PlotState}{vec/matcol/mat, w, h}{The coefficients of our states need to be real numbers before we can plot the
state vector or density operator.  
\\
\\
The procedure will plot a \textbf{vec} object as a bar diagram.  In the
abscissa, all possible kets between $\ket{0,0,..,0}$ and $\ket{(d-1),(d-1),...,(d-1)}$ are labeled from 1 to $d^K$.
For example, 
$$ \left[ \begin {array}{ccc}  1.0&&|0000>
\\\noalign{\medskip} 0.5&&|1100>
\\\noalign{\medskip} 0.25&&|2200>
\end {array} \right] 
$$
\\
with K=2 and d=3, will be displayed as shown in Fig. \ref{pure_plot}:
\\
\begin{figure}[ht]
\centerline{\includegraphics[width=5cm]{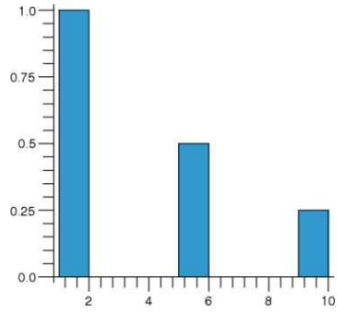}}
\caption{Plot for a pure state (\textbf{vec} object).\label{pure_plot}  }
\end{figure}
\\
To plot a \textbf{matcol} object, the width $w$ and height $h$ of the bars in the bar diagram must be given as
inputs (w=0.5 is the maximum width for the columns not to overlap).  A 3D plot will be the output.  Bras and kets will also be labeled from 1 to $d^K$.  The density matrix for the above state (w=0.5, h=10) is shown in Fig. \ref{mix_plot}:
\\
\begin{figure}[ht]
\centerline{\includegraphics[width=5cm]{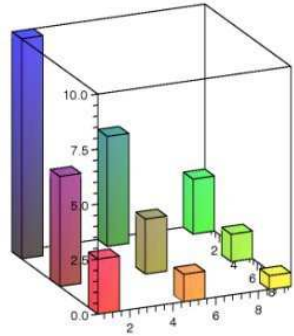}}
\caption{Plot for a density matrix (\textbf{matcol/mat} object).\label{mix_plot}  }
\end{figure}
}{(Matrix, width, height)}{×}{plots, geom3d, Quantavo, LinearAlgebra}
\\
\\
\\
\defi{poly2matcol}{poly}{This procedure transforms a polynomial of the modes\\
$\displaystyle{\mathrm{Poly} = \sum_{m_1,m_2,..,m_K}^{d} \sum_{n_1,n_2,..,n_K}^{d} f(\hat{n},\hat{m}) \prod_{j=1}^{K} a_j^{ n_j} 
 b_j^{ m_j}}$ into an object of type \textbf{matcol}}{polynomial}{Matrix}{Quantavo, LinearAlgebra}
\\
\\
\\
\defi{poly2vec}{poly}{This procedure transforms a polynomial of the modes \\
$\displaystyle{\mathrm{Poly} = \sum_{n_1,n_2,..,n_K}^{d} f(\hat{n}) \prod_{j=1}^{K} a_j^{ n_j} 
 }$ into an object of type \textbf{vec}
 }{polynomial}{Matrix}{Quantavo, LinearAlgebra}
\\
\\
\defi{POVMresult}{matcol, List, vec/matcol}{
If our state before the measurement
is $\ket{\psi}$ or $\rho$, and the POVM elements are described by the set $\{E_m\}$
then the unnormalized state after the measurement will be:\\
POVMresult($E_m$, $List$, $\rho$)
= $\displaystyle{\trace{List}{\left(E_{m List} \otimes \Id_{rest}\right) \; \rho }}$\\
where `$rest$' are all the indexes not included in `$List$'.  
}{Matrix, List, Matrix}{Matrix}{Quantavo, LinearAlgebra}
\\
\\
\defi{Probability}{vec/matcol, List, vec/matcol}{This procedure will calculate the probability of a measurement result or an expected value.
It considers the same cases and objects as the \texttt{Project} procedure.
It uses the definition,
\\
$$P = \frac{\trace{}{E_n \rho}}{\trace{}{\rho}}$$
\\
and assumes $E_n \geq 0$ and $\sum_n E_n = \Id$ to calculate probabilities.
Therefore, one should verify that these conditions hold for the matrices
describing $E_n$ in order to obtain meaningful probabilities.
Below we show more details for different inputs.
\\
\\
\texttt{Probability}($\ket{\psi_1}$, list, $\ket{\psi_2}$)\hskip 1cm $\rightarrow$ \hskip 0.5cm 
$\displaystyle{ \frac{\trace{}{\left(\ket{\psi_1}\bra{\psi_1}_{list} \otimes 1\!\!\:\!{\rm{I}}_{rest}\right) \;\; \ket{\psi_2}\bra{\psi_2}}}{\trace{}{\ket{\psi_2}\bra{\psi_2}}}}$ 
\\
\\
where ($\ket{\psi_1}$, $\ket{\psi_2}$) are converted to \textbf{matcol} objects in an intermediate
step.  One should pay attention to the choice of the projection operator $\ket{\psi_1}\bra{\psi_1}$.  If it is
not nomalized it can give unphysical values for the probability.
\\
\\
Now for a given POVM or Projector,\\
\texttt{Probability}($E_n$, list, $\ket{\psi_2}$) \hskip 1cm $\rightarrow$ \hskip 0.5cm
$\displaystyle{\frac{\trace{}{E_n \ket{\psi_2}\bra{\psi_2}}}{\trace{}{\ket{\psi_2}\bra{\psi_2}}} }$ 
\\
\\
where $E_n$ is assumed to be a \textbf{matcol} object.
\\
\\
Or given a \textbf{vec} and a density operator \textbf{matcol}:\\
\texttt{Probability}($\ket{\psi_1}$, list, $\; \rho\;\;$) \hskip 1cm $\rightarrow$ \hskip 0.5cm
$\displaystyle{\frac{\trace{}{\left(\ket{\psi_1}\bra{\psi_1}_{list} \otimes 1\!\!\:\!{\rm{I}}_{rest}\right) \rho}}{\trace{}{\rho}} }$ 
\\
\\
where $\ket{\psi_1}$ has been converted to a \textbf{matcol} object.\\
\\
\\
And finally for a POVM and a density operator:\\
\\
\\
\texttt{Probability}($E_n$, list, $\;\rho\;\;$) \hskip 1cm $\rightarrow$\hskip 0.5cm 
$\displaystyle{\frac{\mathrm{Tr}\left\{
(E_n \otimes 1\!\!\:\!{\rm{I}}_{rest}) \;\rho\;
\right\}}
{\texttt{Tr}\left\{\; \rho\;\right\}}}$
\\
\\
}{(Matrix, list, Matrix)}{Expression or number}{Quantavo, LinearAlgebra, StateTrace, StateNormalize, vec2matcol}
\\
\\
\defi{Project}{vec/matcol, list, vec/matcol}{
This procedure returns the state after a measurement in the following cases:
\\
\\
\begin{enumerate}
\item if given (\textbf{vec$_1$}, list, \textbf{vec$_2$}) and say \textbf{vec$_1$} and \textbf{vec$_2$} describe respectively $\ket{\psi_1}$ and $\ket{\psi_2}$, \texttt{Project}(\textbf{vec$_1$}, list, \textbf{vec$_2$}) returns the \textbf{vec}
(in principle unnormalized) corresponding to the expresion:
$$\ket{\psi_2'} = \left(\ket{\psi_1}\bra{\psi_1}_{list} \otimes 1\!\!\:\!{\rm{I}}_{rest}\right) \;\; \ket{\psi_2}$$
Note that if for example list=[1,2,3] it means that we want to measure modes one, two an three.  The 
\textbf{vec} object that corresponds to $\ket{\psi_1}$ must therefore have kets with 3 modes.
\\
\\
For example, if $\ket{\psi_2}$ is
\\
 $$V2:=\left[ \begin {array}{ccc} 1&&|0000>
\\\noalign{\medskip}x&&|1100>
\\\noalign{\medskip}{x}^{2}&&|2200>
\end {array} \right] $$
\\
and $\ket{\psi_1}$ is
\\
$$V1:=\left[ \begin {array}{ccc} 1&&|10>
\end {array} \right] $$
\\
Then, \\
\frame{ 
\begin{tabular}{cc} \\ 
&    
\begin{minipage}{4in}  
S := \texttt{Project}(V1, [2,3], V2);
\\
\end{minipage}  
\end{tabular}  
}
\\
\\
will return:
$$S = \left[ \begin {array}{ccc} x &&|1100>
\end {array} \right] $$
\\
\item if given (\textbf{matcol 1}, list, \textbf{vec 2}) and say \textbf{matcol 1} and \textbf{vec 2}
describe $M_n$ and $\ket{\psi_2}$ then it will return the \textbf{matcol} object:
$$ \rho = \left(M_{n_{list}} \otimes 1\!\!\:\!{\rm{I}}_{rest}\right)\; \ket{\psi_2}
\bra{\psi_2} \left(M_{n_{list}} \otimes 1\!\!\:\!{\rm{I}}_{rest}\right)^\dagger $$
\\
So for example, taking the same \texttt{V2} as above and the POVM:
\\
$$M :=  \left[ \begin {array}{ccc} 1&[0,0]&[1,1]\\\noalign{\medskip}1&[1,1]&[0
,0]\\\noalign{\medskip}1&[1,1]&[1,1]\end {array} \right] $$
\\
Then, \\
\frame{ 
\begin{tabular}{cc} \\ 
&    
\begin{minipage}{4in}  
S := \texttt{Project}(M, [1,2], V2);
\\
\end{minipage}  
\end{tabular}  
}
\\
\\
will return the state:
$$S =  \left[ \begin {array}{ccc} x \bar{x}  &[0,0,0,0
]&[0,0,0,0]\\\noalign{\medskip}x+x \bar{x} &[0
,0,0,0]&[1,1,0,0]\\\noalign{\medskip}x\bar{x} +\bar{x} &[1,1,0,0]&[0,0,0,0]
\\\noalign{\medskip}x+x\bar{x} +\bar{x}+1&[1,1,0,0]&[1,1,0,0]\end {array}
 \right] 
 $$
 which is the mixed state that corresponds to:
$$ S =  \left[ \begin {array}{cc} x&[0,0,0,0]\\\noalign{\medskip}x+1&[1,1,0,0
]\end {array} \right] $$
\\
\item if given (\textbf{vec 1}, list, \textbf{matcol 2}) and say \textbf{vec 1}, \textbf{matcol 2} 
describe $\ket{\psi_1}$ and $\rho$ respectively, then it will build $\ket{\psi_1}\bra{\psi_1}$ and return the 
\textbf{matcol} object corresponding to:
$$\rho' = \left(\ket{\psi_1}\bra{\psi_1}_{list} \otimes 1\!\!\:\!{\rm{I}}_{rest} \right)\; \rho \;  \left(\ket{\psi_1}\bra{\psi_1}_{list} \otimes 1\!\!\:\!{\rm{I}}_{rest} \right)$$
\\
\\
\item if given (\textbf{matcol 1}, list, \textbf{matcol 2}) and say \textbf{matcol 1} and \textbf{matcol 2}  describe $M_n$ and $\rho$ respectively, then it will return:
$$\rho' =  \left(M_{n_{list}} \otimes 1\!\!\:\!{\rm{I}}_{rest}\right) \; \rho \;  \left(M_{n_{list}} \otimes 1\!\!\:\!{\rm{I}}_{rest}\right)^+$$
\end{enumerate}
In a nutshell:
\\
\\
\begin{tabular}{lcl}
\texttt{Project}($\ket{\psi_1}$, list, $\ket{\psi_2}$)& $\rightarrow$& 
$\left(\ket{\psi_1}\bra{\psi_1}_{list} \otimes 1\!\!\:\!{\rm{I}}_{rest}\right) \;\; \ket{\psi_2}$
\\
\texttt{Project}($M_n$, list, $\ket{\psi_2}$) & $\rightarrow$& $  \left(M_{n_{list}} \otimes 1\!\!\:\!{\rm{I}}_{rest}\right)\; \ket{\psi_2}
\bra{\psi_2} \left(M_{n_{list}} \otimes 1\!\!\:\!{\rm{I}}_{rest}\right)^\dagger $
\\
\texttt{Project}($\ket{\psi_1}$, list,$\;\; \rho\;\;$) & $\rightarrow$& $\left(\ket{\psi_1}\bra{\psi_1}_{list} \otimes 1\!\!\:\!{\rm{I}}_{rest} \right) \; \rho \;  \left(\ket{\psi_1}\bra{\psi_1}_{list} \otimes 1\!\!\:\!{\rm{I}}_{rest} \right)$
\\
\texttt{Project}($M_n$, list,$\;\;\rho\;\;$) & $\rightarrow$& $ \left(M_{n_{list}} \otimes 1\!\!\:\!{\rm{I}}_{rest}\right) \; \rho \;  \left(M_{n_{list}} \otimes 1\!\!\:\!{\rm{I}}_{rest}\right)^+$\\\\
\end{tabular}
}{(Matrix, list Matrix)}{Matrix}{Quantavo, StateComplexConjugate, LinearAlgebra, vec2matcol}
\\
\\
\defi{PS}{vec/matcol, j, $\phi$}{PHASE SHIFTER:
This procedure makes a phase shifter transformation to our state. It is implemented
making the following mode transformation to the specified mode $j$:  
\begin{eqnarray}
a_j' = e^{i \phi} a_j  \label{one} \\
b_j' = e^{-i \phi} b_j \label{two}
\end{eqnarray}
(that is transformation (\ref{one}) for a \textbf{vec} and transformation (\ref{one}) and  (\ref{two}) for
\textbf{matcol} and \textbf{mat}.)
}{(Matrix, whole number, symbol or number)}{Matrix}{Quantavo, LinearAlgebra}
\\
\\
\begin{LARGE}S\end{LARGE}\\
\\
\defi{SqueezedVac}{m,d,$\lambda$}{ This procedure builds a \textbf{vec} describing the state
 $\ket{\phi} \sim \sum_{n=0}^{d-1} \lambda^n \ket{n}^{\otimes m}$ where $m$ can be $m=1,2$. 
 The state is not normalized.}{(whole number, whole number, string or number)}{2 column Matrix}{Quantavo, LinearAlgebra}
\\
\\
\defi{StateApprox}{vec/matcol, list, n}{This procedure is used to reduce the size and complexity of the \textbf{vec} or \textbf{matcol} objects with an approximation.\\\\
Numerical Approximation:  If the coefficients of our state are numbers then use as follows:
\texttt{StateApprox}(vec/matcol, [], n).  All Rows for which the coefficient $< 10^{-n}$ will be deleted.\\\\
Symbolic Approximation:  If the coefficients are symbolic polynomials (or monomials) and some
variables are small, we may choose to delete all terms with a certain power in those variables.  If we execute
S := StateApprox(M,[$lambda$,$r$],10), then all terms containing $\lambda^n r^m$ such that $n+m > 10$ will 
be deleted. For example if we have a state of the form:
 $$M := \left[ \begin {array}{cc} 1+{y}^{4}{x}^{2}+y{x}^{4}&[0,0,0,0]
\\\noalign{\medskip}{y}^{7}x+{x}^{3}+{x}^{5}&[1,1,0,0]
\\\noalign{\medskip}{x}^{2}+{x}^{4}+{x}^{6}+{y}^{5}&[2,2,0,0]
\end {array} \right] $$
then, \\
\frame{ 
\begin{tabular}{cc} \\ 
&    
\begin{minipage}{4in}  
S := StateApprox(M,[y,x],5);
\\
\end{minipage}  
\end{tabular}  
}
\\
\\
will return the state:
$$ \left[ \begin {array}{cc} 1+y{x}^{4}&[0,0,0,0]\\\noalign{\medskip}{x}
^{3}+{x}^{5}&[1,1,0,0]\\\noalign{\medskip}{x}^{2}+{x}^{4}+{y}^{5}&[2,2
,0,0]\end {array} \right] 
$$\\
\\
and\\
\frame{ 
\begin{tabular}{cc} \\ 
&    
\begin{minipage}{4in}  
S := StateApprox(M,[y,x],1);
\\
\end{minipage}  
\end{tabular}  
}
\\
\\
will return the state:
 $$\left[ \begin {array}{cc} 1&[0,0,0,0]\end {array} \right] $$}
 {(Matrix, list, integer)}{Matrix}{Quantavo, LinearAlgebra} 
\\
\\
\defi{StateComplexConjugate}{matcol/mat}{
It will complex conjugate all the coefficients of an object of type \textbf{mat} or \textbf{matcol}.
}{Matrix}{Matrix}{Quantavo, LinearAlgebra}
\\
\\
\defi{StateMultiply}{number/matcol, matcol/vec}
{This procedure multiplies following the matrix multiplication rules the following objects:\\\\
k $\times$ \textbf{matcol},\\
k  $\times$ \textbf{vec}, \\ 
\textbf{matcol} $\times$ \textbf{matcol}\\
\textbf{matcol}  $\times$ \textbf{vec}\\\\
where k is a number, a variable, a function or a string\\}
{(string/Matrix, Matrix)} 
{Matrix} 
{Quantavo, LinearAlgebra}
\\
\\
\defi{StateSort}{vec/matcol}
{This procedure sorts by rows the objects \textbf{vec} or \textbf{matcol}.  The order is determined
in the following way.  The modes are converted from the numeral base $d$ to the numeral base 10 and are
then sorted by increasing order.  For \textbf{matcol} they are sorted first according to the order of the ``kets'' and inside the same ``ket'' number, by the order of the ``bras''.  For example, with 
$d =2$ and $K=2$,
$[\lambda, [0,0] ] < [\lambda^2, [1,0] ] < [\lambda, [0,1] ] < [\lambda^3, [1,1] ]$.  The output is the ordered state. }
{Matrix} 
{Matrix} 
{Quantavo, LinearAlgebra, Tribullesmatcol, Tribullesvec}
\\
\\
\defi{StateTrace}{mat/matcol}{This procedure performs the trace on \textbf{mat} or \textbf{matcol} states. It simply
adds all diagonal elements.}{Matrix}{Real Number or symbolic expression}{Quantavo, LinearAlgebra}
\\
\\
\defi{StateNorm}{vec}{It evaluates the norm $\langle \phi |\phi \rangle$ of the state associated with the object \textbf{vec}}{Matrix}{number or expression}{Quantavo, LinearAlgebra}
\\
\\
\defi{StateNormalize}{vec/matcol/mat}{This procedure returns a normalized state such that
$\langle \psi | \psi \rangle = 1 $ for \textbf{vec} and Tr$(\rho)$=1 for \textbf{matcol} or \textbf{mat}. 
If the state was not normalized to begin with it will print:  ``the state is not normalized'' and
then return the normalized state.  }{Matrix}{Matrix and printed answer}{Quantavo, LinearAlgebra, IsNormalized, LittleTrace}
\\
\\
\defi{StatePartialTranspose}{vec/matcol/mat,s}{This procedure works only for bipartite states (two modes only).  If your state is not bipartite use \texttt{Traceout} to trace out the other modes.  The parameter `\textbf{s}' 
specifies if the partial transposition must be made with respect to the first or to the second mode.  The output is
the partially transposed \textbf{matcol} or \textbf{mat}.}{(Matrix, 1 or 2)}{Matrix}{Quantavo, LinearAlgebra}
\\
\\
\begin{LARGE}T\end{LARGE}\\
\\
\defi{TensorProduct}{vec/matcol, List, vec/matcol, List}{
This will return the tensor product between modes [$List_A$] of state A and modes [$List_B$] of state B.  
It therefore implements the operation,
($\rho$,listA,$\sigma$, listB)  $\longrightarrow \rho_{listA} \otimes \sigma_{listB}$.  Note
that the dimension of each state and its corresponding list must be equal.
So for example we can declare two states,
\\
\frame{
\begin{tabular}{cc} \\
&   
\begin{minipage}{4in}
Sq:=\texttt{SqueezedVac}(2, 3, $\xi$);\\
Fock2:=Matrix([[$\gamma$, [1,1]], [$gamma$, [2,2]]]);
\\
\end{minipage}
\end{tabular}
}
\\
\\
\begin{tabular}{cc}
$\left[ \begin{array}{cc} 1&[0,0]\\\noalign{\medskip}\xi&[1,1]
\\\noalign{\medskip}{\xi}^{2}&[2,2]\end {array} \right] $
&
$\left[ \begin{array}{cc} \gamma&[1,1]\\\noalign{\medskip}\gamma&[2,2
]\end {array} \right] $\\
\end{tabular}
\\
\\
and after using 
\\
\frame{
\begin{tabular}{cc} \\
&   
\begin{minipage}{4in}
TensorProduct(Sq,[1,4],Fock2,[2,3]);
\\
\end{minipage}
\end{tabular}
}\\
we obtained the tensored \textbf{vec} object,
$$\left[ \begin {array}{cc} \gamma&[0,1,1,0]\\\noalign{\medskip}\gamma&
[0,2,2,0]\\\noalign{\medskip}\xi\,\gamma&[1,1,1,1]\\\noalign{\medskip}
\xi\,\gamma&[1,2,2,1]\\\noalign{\medskip}{\xi}^{2}\gamma&[2,1,1,2]
\\\noalign{\medskip}{\xi}^{2}\gamma&[2,2,2,2]\end {array} \right] 
$$
}{Matrix,List,Matrix,List}{Matrix}{Quantavo, LinearAlgebra}
\\
\\
\defi{TensorVac}{vec/matcol/mat, m}{This procedure tensors `m' vacuum modes with an existing state.  The new modes are in a product state with the original state.  If it is a \textbf{vec} object it will therefore do the transformation $\ket{\psi} \longrightarrow \ket{\psi} \otimes \ket{0}^{\otimes m}$ and
$\rho \longrightarrow \rho \otimes (\ket{0}\bra{0})^{\otimes m}$ for \textbf{mat} and \textbf{matcol}.  The procedure will also transform the global variable $\textbf{K} \rightarrow \textbf{K}+\textbf{m}$.}{(Matrix, whole number)}{Matrix}{Quantavo, LinearAlgebra}
\\
\\
\defi{Traceout}{matcol,i}{This procedure takes the partial trace of a sparse density matrix with respect to mode $i$.  Its input is an object of type \textbf{matcol}}{(Matrix, whole number)}{Matrix}{ Quantavo, LinearAlgebra}
\\
\\
\defi{Trim}{Vector/vec/matcol}{This procedure eliminates all the non zero entries of a \textbf{vector} of dimension $d^K$ and converts it into an object of type \textbf{vec}.  It also deletes all non-zero entries in
objects of type \textbf{vec} and \textbf{matcol}}{Vector/Matrix}{Matrix}{Quantavo, LinearAlgebra}
\\
\\
\begin{LARGE}U\end{LARGE}\\
\\
\defi{UnitaryEvolution}{Unitary Matrix, vec/matcol}{ 
Whether we have just built a unitary matrix with \texttt{BuildUnitary} or we have an arbitrary unitary matrix, we can use this procedure as follows:
\\
\\
\frame{ 
\begin{tabular}{cc} \\ 
&    
\begin{minipage}{4in}  
V:= \texttt{UnitaryEvolution}(U, \textbf{vec/matcol}): 
\\
\end{minipage}  
\end{tabular}  
}
\\
\\
where U is the unitary matrix of dimension $K \times K$ and our state is described by a
\textbf{vec} or a \textbf{matcol} object.  This will effectively implement the mode transformation:
$$ \overline{a}' = U \overline{a}$$
and return the \textbf{vec} or \textbf{matcol} after the transformation. (Note that for \textbf{matcol}
it also does $ \overline{b}' = U^\dagger \overline{b}$ )
}{(Matrix, Matrix)}{Matrix}{Quantavo, LinearAlgebra}
\\
\\
\begin{LARGE}V\end{LARGE}\\
\\
\defi{Vac}{Nr. of modes}{This procedure returns a vacuum \textbf{vec} with the number of specified modes.  
For example \texttt{Vac}(3) will return  $\left[ \begin {array}{cc} 1&[0,0,0]\end {array} \right]$ 
}{positive natural number}{Matrix}{Quantavo, LinearAlgebra}
\\
\\
\defi{vec2mat}{vec}{It converts a \textbf{vec} object into a \textbf{mat} object.
The procedures is independent of $d$ and $K$.}{2 column Matrix}{Matrix}{Quantavo, LinearAlgebra}
\\
\\
\defi{vec2matcol}{vec}{It converts a \textbf{vec} object into a \textbf{matcol} object effectively
doing 
$$\sum_{\bar{n}} \alpha_{\bar{n}} \ket{\bar{n}} \longrightarrow  
\sum_{\bar{n}\bar{m}} \alpha_{\bar{n}} \bar{\alpha}_{\bar{m}} \ket{\bar{n}}\bra{\bar{m}}$$
Note that no normalization is implemented.  The procedure is independent of $d$ and $K$.}{2 column Matrix}{3 Column Matrix}{Quantavo, LinearAlgebra}
\\
\\
\defi{vec2poly}{vec}{It converts a \textbf{vec} object into a \textbf{poly} object}{2 column Matrix}{Polynomial}{LinearAlgebra, Quantavo}
\\
\\
\defi{VectorModes}{i}{this procedure takes as input an integer between 1 and $d^K$ and outputs a list with the equivalent number of photons in each mode.  This way, \texttt{VectorModes}(1) = [0,0,0] if it is a 3 mode state and \texttt{VectorModes}(2)=[1,0,0] if $K=3$ and $d=2$ for example.  (Note that $d$ and $K$ need to be defined)}{integer}{List}{Quantavo}
\\
\\
\defi{VectorRow}{Indi, d}{$d$ (maximum number of photons) and $K$ (number of modes) need to be specified.  This is the inverse procedure of \texttt{VectorModes}.  When a list with the
number of photons in each mode (for instance Indi = [0,1,0]) and the value of $d$ are given, it outputs the row number that corresponds to it in a \textbf{vector} type object).  This way, for $d=2$, $K=3$ VectorRow([0,1,0],2)=3.}{(List,d)}{whole number}{Quantavo}
\\
\\
Quantavo also uses the local procedures:
\\
\\
multiplymatcol, multiplymatcolvec, Tribullesmatcol, Tribullesvec, VectorModes, VectorRow, indexvec, 
modesvec, vecBS, myvecBS, matcolBS, mymatcolBS, Projectvecvec, Projectmatcol, Dvec, Dmat, Dmatcol,
barra,histo;

\section{Copyright and Disclaimer}

Copyright (c) 2008 Alvaro Feito Boirac.\\\\
This is the Module QUANTAVO, a toolbox for Quantum Optics calculations 
that can be used in Maple$^{TM}$ (Waterloo Maple Inc.)\\
\\
It is released under the GNU General Public License v3 which can be obtained at
\url{http://www.gnu.org/licenses/gpl.html} .
Please acknowledge its use if used to establish results for a published work. If you make any improvements or find any bugs the author will be thankful if you can let him know.\\

\textbf{DISCLAIMER}: This Module is distributed in the hope that it will be useful, but WITHOUT ANY WARRANTY, without even the implied warranty of MERCHANTABILITY or FITNESS FOR A PARTICULAR PURPOSE.\\\\ 
The author can be contacted at \href{mailto:ab1805@imperial.ac.uk}{ab1805@imperial.ac.uk}

\end{document}